\documentclass{article}
\usepackage{spconf,amsmath,graphicx}
\usepackage{booktabs}
\usepackage{lipsum}
\def\x{{\mathbf x}}
\def\L{{\cal L}}

\usepackage{hyperref}
\usepackage{url}
\usepackage{amsmath,graphicx}
\usepackage{amssymb,multirow}
\usepackage{caption}
\usepackage{subcaption}

\usepackage{verbatim}
\usepackage{bm, mathtools} 

\usepackage{url}
\usepackage{multibib}
\newcites{app}{Appendix Reference}

\title{End-to-End Text-to-Speech using Latent Duration based on VQ-VAE }

\name{Yusuke Yasuda$^{\dagger\star}$, Xin Wang$^\dagger$, Junichi Yamagishi$^{\dagger\star}$\thanks{This work was partially supported by a JST CREST Grant (JPMJCR18A6, VoicePersonae project), Japan,  MEXT KAKENHI Grants (16H06302, 18H04120, 18H04112, 18KT0051, 19K24371), Japan.  The numerical calculations were carried out on the TSUBAME 3.0 supercomputer at the Tokyo Institute of Technology.}}
\address{$^\dagger$National Institute of Informatics, Japan 
$^\star$The Graduate University for Advanced Sciences, Japan 
}
\begin{document}
\ninept

\maketitle

\begin{abstract}
Explicit duration modeling is a key to achieving robust and efficient alignment in text-to-speech synthesis (TTS). We propose a new TTS framework using explicit duration modeling that incorporates duration as a discrete latent variable to TTS and enables joint optimization of whole modules from scratch. We formulate our method based on conditional VQ-VAE to handle discrete duration in a variational autoencoder and provide a theoretical explanation to justify our method. In our framework, a connectionist temporal classification (CTC) -based force aligner acts as the approximate posterior, and text-to-duration works as the prior in the variational autoencoder. We evaluated our proposed method with a listening test and compared it with other TTS methods based on soft-attention or explicit duration modeling. The results showed that our systems rated between soft-attention-based methods (Transformer-TTS, Tacotron2) and explicit duration modeling-based methods (Fastspeech).
\end{abstract}


\section{Introduction}
\label{sec:intro}
\vspace{-5pt}

Sequence-to-sequence text-to-speech (TTS) typically consists of a text encoder to encode an input character or phoneme sequence, an acoustic decoder to generate an acoustic feature sequence, and a neural vocoder to convert the acoustic features into output waveforms. Another indispensable part is the learning of alignment between input and output sequences. 
Most of the recent models \cite{Wang2017, Sotelo2017Char2wavES, DBLP:conf/iclr/PingPGAKNRM18, Shen2017, DBLP:conf/aaai/Li0LZL19} use soft-attention \cite{Bahdanau2014,DBLP:conf/nips/ChorowskiBSCB15} and learn the ``probabilities'' for each output step of being aligned with input steps. There is also a TTS model using hard-attention \cite{Yasuda2019} and one that learns the latent alignment with a forward-backward algorithm \cite{rabiner1989tutorial}.

Being either deterministic or latent, both soft- and hard-attention techniques parameterize the alignment on the basis of an atomic probability $P(\alpha_n=m)$ that indicates the likelihood of the $n$-th output step being aligned with the $m$-th input step. Such a probability, however, does not directly tell how many output steps are likely to be emitted from the $m$-th input token. For TTS tasks, knowing the probability $P(l_m = k)$ of generating $k\in\mathbb{N}$ output steps from the $m$-th input token is more useful from the perspective of speech production process. It also allows direct control of the duration $l$, avoids the engineering trick in soft-attention to predict ``when to stop'', and guarantees monotonic alignment between input and output.

In this paper, we propose a new neural TTS system with such a latent duration model. In contrast to soft- or hard-attention-based systems, our proposed latent duration model directly parameterizes $P(l_m = k)$, as shown in Fig.~\ref{fig:alignment}. 
Moreover, in contrast to recent neural TTS systems that also replace attention with duration models \cite{Ren2019,DBLP:conf/icassp/ZengWCXX20,DBLP:journals/corr/abs-1909-01700,DBLP:journals/corr/abs-2005-07799,DBLP:journals/corr/abs-2006-04558}, our latent duration model is jointly trained with other components in the TTS systems. This is achieved using a deep variational Bayesian approach \cite{DBLP:journals/corr/KingmaW13}. Since the duration in TTS systems is discrete, our proposed model is similar to the vector quantized auto-encoder (VQ-VAE) \cite{DBLP:conf/nips/OordVK17}, but rather than simply following the heuristic VQ-VAE training criteria, we define our VQ-VAE framework on the basis of variational inference theory. 
Experiments on an English speech database showed promising results when comparing our proposed model with Tactron \cite{Shen2017}, Transformer \cite{DBLP:conf/aaai/Li0LZL19}, and Fast Speech TTS systems \cite{Ren2019}. 

After introducing related methods in Section~\ref{sec:related-works}, we describe the proposed TTS system in Section \ref{sec:proposed-method}. Experiments are explained in Section \ref{sec:experiments}. We conclude with a brief summary in Section \ref{sec:conclusion}.

\section{Related works}
\label{sec:related-works}
\vspace{-5pt}

\subsection{Duration modeling in TTS}
\vspace{-3pt}

Explicit duration modeling was recently incorporated into neural TTS systems as an alternative to soft- and hard-attention for more robust alignment modeling.
For example, FastSpeech \cite{Ren2019} trains a duration predictor with alignment obtained from a soft-attention-based teacher model, 
AlignTTS \cite{DBLP:conf/icassp/ZengWCXX20} extracts duration using an aligner based on hard and monotonic alignment, and DurIAN \cite{DBLP:journals/corr/abs-2005-07799} and FastSpeech2 \cite{DBLP:journals/corr/abs-2006-04558} use external force aligners to extract the duration.

All four systems above use independent duration and aligner models, and as such they require multiple training phases to construct a whole TTS system. 
In contrast, recent models such as JDI-T \cite{DBLP:journals/corr/abs-2005-07799} and Glow-TTS \cite{DBLP:journals/corr/abs-2005-11129} can train the TTS and duration model jointly. Glow-TTS \cite{DBLP:journals/corr/abs-2005-11129} treats duration as a latent variable, although it is not formulated as a VAE.
Note that all methods above treat duration as a continuous value, even though duration in those TTS systems corresponds to the number of frames, which is intrinsically discrete.

Our proposed method differs from the previous works in three key aspects: 1) duration is modeled as a discrete variable,
2) duration is treated as a latent variable using the VQ-VAE framework, and 3) all components in our model are jointly trainable from scratch.
Note that duration has also been treated as a latent variable in the classical hidden semi-Markov based parametric speech synthesis \cite{DBLP:conf/interspeech/ZenTMKK04}.

\subsection{Vector quantized autoencoder for speech tasks}
\label{ssec:vq-vae}
\vspace{-3pt}

VQ-VAE \cite{DBLP:conf/nips/OordVK17} has been applied to various speech synthesis tasks, including diverse and controllable TTS \cite{henter2018deep, Sun2020}, a new TTS framework based on symbol-to-symbol translation \cite{DBLP:journals/corr/abs-2005-05525}, speech coding \cite{DBLP:conf/icassp/GarbaceaOLLLVW19}, voice conversion \cite{Ding2019}, and representation learning \cite{DBLP:journals/taslp/ChorowskiWBO19, Tjandra2019, Wang2020, DBLP:journals/corr/abs-2005-07884}.

Among the related methods, Sun et al. applied conditional VQ-VAE to TTS, although their objective was diverse TTS for data augmentation rather than duration modeling \cite{Sun2020}. 
Their method relies on soft-attention to align speech and phoneme.
The new TTS framework using VQ-VAE \cite{DBLP:journals/corr/abs-2005-05525} treats TTS as sequence transduction by encoding target speech into a sequence of discrete phoneme-like symbols. As a result, mapping from text to speech symbols can be tackled using machine translation methods. However, they still require soft-attention to align text and speech symbols.

Our system uses VQ-VAE to model the latent discrete duration, which has never been explored before. Furthermore, we derive the training criteria on the basis of variational inference.

\section{Proposed TTS system}
\label{sec:proposed-method}
\vspace{-5pt}
In this section, we explain the proposed TTS system as a probabilistic model and then describe its implementation.

\subsection{Model definition and variational lower bound}
\label{sec:model_def}
Let us write the input sequence of discrete tokens as $\mathbf{y}_{1:U}=(y_{1}, \cdots  y_U)$, where $y_u$ is the $u$-th token (e.g., letter or phone). 
We then use $\mathbf{x}_{1:T}=(\mathbf{x}_1, \cdots, \mathbf{x}_T)$ to denote an output sequence of acoustic features, where $\mathbf{x}_t \in \mathbb{R}^{O}$ is for the $t$-th output time step or \emph{frame}. Our goal is to learn a good model $p(\mathbf{x}_{1:T} \mid \mathbf{y}_{1:U})$.

\begin{figure}[!t]
\centering
\includegraphics[trim=5 520 75 270,clip,width=1.0\columnwidth]{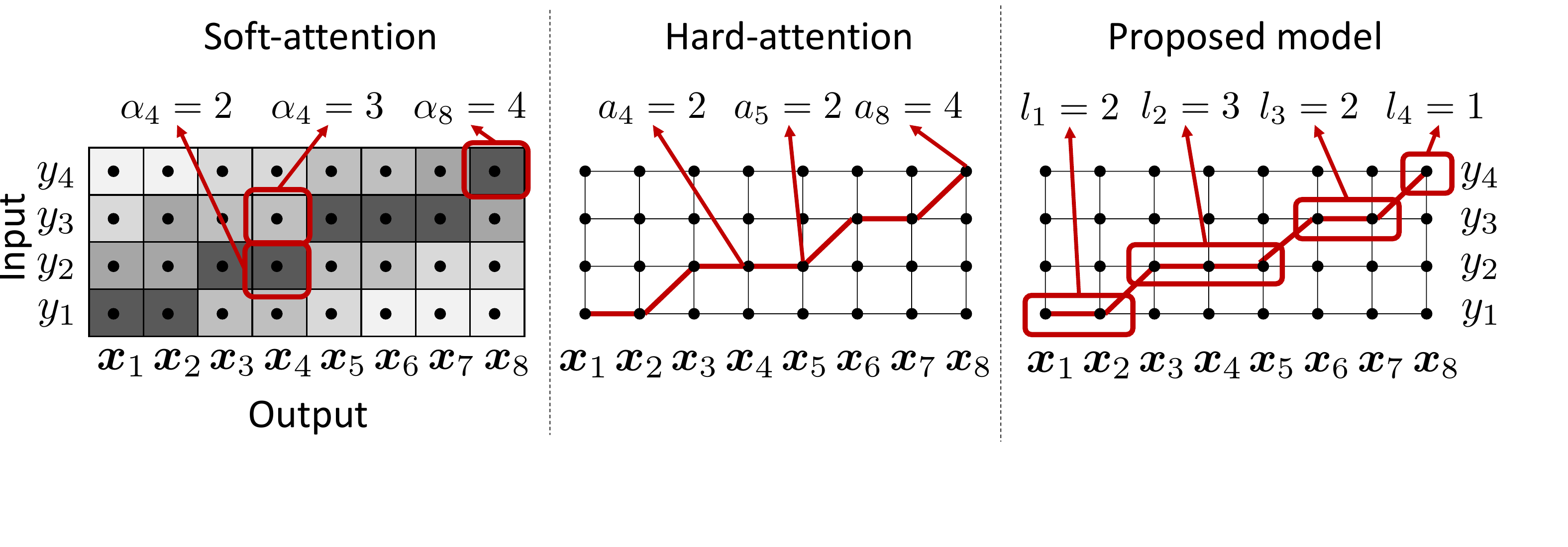}
\vspace{-8mm}
\caption{Illustration of alignment in soft-attention (left), hard-attention (middle), and proposed models (right). In soft- and hard-attention, alignments $\alpha_{n}=m$ and $a_{n}=m$ denote that output $\mathbf{y}_n$ is aligned with input $x_m$. In the proposed model, $z_m=l$ indicates that $x_m$ emits $l\in\mathbb{N}$ consecutive output steps.}
\label{fig:alignment}
\end{figure}

We use $\mathbf{l}_{1:U}=(l_1, \cdots, l_U)$ to denote the latent duration for $\mathbf{y}_{1:U}$, where the value of $l_u$ is equal to the duration of the $u$-th input token. For example, in Fig.~\ref{fig:alignment}, $l_2 = 3$ indicates that the second input token $y_2$ is aligned with three output frames $(\mathbf{x}_3,\mathbf{x}_4,\mathbf{x}_5)$. For an output sequence with a total duration of $T$, we constrain $T = \sum_u l_{u}$. We further constrain $1 \leq l_u \leq K$, where $K$ is a hyper-parameter that decides the maximum value of duration. 
 
Since $l_u$ is discrete for TTS systems, we can parameterize $p(\mathbf{x}_{1:T} \mid \mathbf{y}_{1:U})$ with VQ-VAE. Let $\mathcal{Z} = \{\mathbf{e}_1, \cdots, \mathbf{e}_K\}$ be a codebook with $K$ code words, where $\mathbf{e}_k \in \mathbb{R}^{D}$. We then introduce a quantized latent vector $z^{(q)}_{u}\in{\mathcal{Z}}$ and an un-quantized latent vector $z^{(r)}_{u} \in \mathbb{R}^{D}$ for the $u$-th input token. 
Because $z^{(q)}_{u}\in{\mathcal{Z}}$, we can link $z^{(q)}_{u}$ with $l_u$ by setting $z^{(q)}_{u} = \mathbf{e}_{l_u}, \forall{u}\in\{1,\cdots,U\}$, which avoids modeling $z^{(q)}_{u}$ and $l_u$ independently.
We additionally introduce $z^{(r)}_{u}$ because it allows us to derive the VQ-VAE training criteria in a theoretically sound manner \cite{henter2018deep}.

Model training through direct optimization of $p(\mathbf{x}_{1:T}|\mathbf{y}_{1:U})$ is daunting, but the above definition allows us to find a tractable variational lower bound (ELBO)\footnote{Details can be found in the appendix on \cite{yasuda2020endtoend}.}:
\begin{align}
&\log p(\mathbf{x}_{1:T} \mid \mathbf{y}_{1:U}) 
\nonumber\\
= & \log\sum_{\forall \mathbf{z}^{(q)}_{1:U}}\int_{\mathbf{{z}}^{(r)}_{1:U}} p(\mathbf{x}_{1:T}, \mathbf{z}^{(q)}_{1:U}, \mathbf{{z}}^{(r)}_{1:U} \mid \mathbf{y}_{1:U}) d\mathbf{{z}}^{(r)}_{1:U} \label{eq:model_exact}\\
\ge &\mathbb{E}_{Q_{\lambda}(\mathbf{z}^{(q)}_{1:U})}[\underbrace{\log p_\theta(\mathbf{x}_{1:T} \mid \mathbf{z}^{(q)}_{1:U}, \mathbf{y}_{1:U})}_{\text{Decoder}}] \label{eq:elbo_decoder}\\
~ &-\mathrm{KL}[{Q_{\lambda}(\mathbf{z}^{(q)}_{1:U}}) \| \underbrace{P_\phi(\mathbf{z}^{(q)}_{1:U} \mid \mathbf{y}_{1:U})}_{\text{Prior}}] \label{eq:elbo_prior}\\ 
~ &- \mathbb{E}_{Q_{\lambda}(\mathbf{z}^{(q)}_{1:U})} \Big\{\mathrm{KL}[{Q_{\psi}(\mathbf{{z}}^{(r)}_{1:U} \mid \mathbf{z}^{(q)}_{1:U})} \| \underbrace{p(\mathbf{{z}}^{(r)}_{1:U} \mid \mathbf{z}^{(q)}_{1:U})}_{\text{Vector quantization}}] \Big\} \label{eq:elbo_vq}.
\end{align}
To avoid notation clutter, we hide the condition $\mathbf{x}_{1:T}, \mathbf{y}_{1:U}$ in approximate posteriors $Q_{\psi}(\mathbf{z}^{(r)}_{1:U} \mid \mathbf{z}^{(q)}_{1:U}, \mathbf{y}_{1:U}, \mathbf{x}_{1:T})$ and $Q_{\lambda}(\mathbf{z}^{(q)}_{1:U} \mid \mathbf{y}_{1:U}, \mathbf{x}_{1:T})$. $\mathbf{x}_{1:T}$ in the decoder is assumed to be independent of $\mathbf{z}^{(r)}_{1:U}$ given $\mathbf{z}^{(q)}_{1:U}$ and $\mathbf{y}_{1:U}$. Furthermore, $\mathbf{z}^{(r)}_{1:U}$ is assumed to   depend only on $\mathbf{z}^{(q)}_{1:U}$.
$\text{KL}$ denotes the 
Kullback–Leibler divergence.

\subsection{Model parameterization}
\label{sec:model_parameterization}
Each term in the ELBO of Eq.~(\ref{eq:elbo_decoder}--\ref{eq:elbo_vq}) can be parameterized using neural networks and calculated in a closed form.

\subsubsection{Approximate posteriors}
\label{sec:app_posterior}
We first define the approximate posterior for $\mathbf{z}^{(q)}_{1:U}$ as
\begin{align}
&Q_{\lambda}(\mathbf{z}^{(q)}_{1:U} \mid \mathbf{y}_{1:U}, \mathbf{x}_{1:T}) \nonumber\\
= & \prod_{u=1}^U P(\mathbf{z}^{(q)}_u \mid \mathbf{y}_{1:U}, \mathbf{x}_{1:T}) = \prod_{u=1}^U \mathcal{I}(\mathbf{z}^{(q)}_u = \mathbf{e}_{l_u}), \label{eq:indicator}
\end{align}
where  $\mathcal{I}(\cdot)$ is an indicator function and $l_u$ is the code word index (and duration) for the $u$-th input token. We then define 
\begin{align}
&Q_{\psi}(\mathbf{z}^{(r)}_{1:U} \mid \mathbf{z}^{(q)}_{1:U}, \mathbf{y}_{1:U}, \mathbf{x}_{1:T}) \nonumber\\
=& \prod_{u=1}^U p(\mathbf{z}^{(r)}_u \mid  \mathbf{z}^{(q)}_{u-1}, \mathbf{y}_{1:U}, \mathbf{x}_{1:T}) = \prod_{u=1}^U \mathcal{N}(\mathbf{z}^{(r)}_{u}; \mathbf{d}_u, \sigma^2\mathbf{I}), \label{eq:f_uni} \end{align}
where $\mathcal{N}$ is a multivariate Gaussian,  $\sigma^2$ is a hyper-parameter,
and $\mathbf{d}_u$ is computed by a neural network as $\mathbf{d}_{u} =\mathrm{LatentNet}_\psi(\mathbf{z}^{(q)}_{u-1}, \mathbf{\bar{x}}_u, y_u)$. 

Since $\mathbf{x}_{1:T}$ has $T$ frames while $\mathbf{y}_{1:U}$ has $U$ tokens, we aggregate $\mathbf{x}_{1:T}$ as $\mathbf{\bar{x}}_{1:U} = (\mathbf{\bar{x}}_1, \cdots, \mathbf{\bar{x}}_U)$ before feeding it to $\mathrm{LatentNet}_\psi(\cdot)$. The aggregation is conducted by averaging $\mathbf{x}_t$ that correspond to the $u$-th token, given the value of $\mathbf{l}_{1:u}$. For example, in Fig.~\ref{fig:alignment}, we get $\mathbf{\bar{x}}_1 = \frac{\mathbf{x}_1 + \mathbf{x}_2}{2}$ and $\mathbf{\bar{x}}_2 = \frac{\mathbf{x}_3 + \mathbf{x}_4 + \mathbf{x}_5}{3}$ given $l_1=2$ and $l_2 = 3$. 

We let $Q_{\lambda}$ be an indicator function, following the idea of VQ-VAE. The value of $l_u$ during training is produced by another surrogate model (explained in Section~\ref{sec:ctc}).
A Gaussian posterior $Q_{\psi}$ is inspired by the original VAE \cite{DBLP:journals/corr/KingmaW13}. It is used to compute the KL divergence in Eq.~(\ref{eq:elbo_vq}), which is explained in  Section~\ref{sec:vq}.

\subsubsection{Decoder}
\label{sec:decoder}
Similar to other models, our model uses an auto-regressive decoder. 
Given $\mathbf{y}_{1:U}$, $\mathbf{l}_{1:U}=(l_1, \cdots, l_U)$, and $\mathbf{z}^{(q)}_{1:U}=(\mathbf{z}^{(q)}_{1}=\mathbf{e}_{l_1}, \cdots, \mathbf{z}^{(q)}_{U}=\mathbf{e}_{l_U})$, the PDF of  $\mathbf{x}_{1:T}$ is defined as
\begin{align}
&p_\theta(\mathbf{x}_{1:T} \mid \mathbf{z}^{(q)}_{1:U}, \mathbf{y}_{1:U}) \nonumber \\
=& \prod_{t=1}^T p(\mathbf{x}_t\mid \mathbf{x}_{t-1}, \mathbf{\widehat{z}}^{(q)}_t, \widehat{y}_t) = \prod_{t=1}^T \mathcal{N}(\mathbf{x}_t; \mathbf{\mu}_t, \sigma_d^2\mathbf{I}), \label{eq:decoder}
\end{align}
where $\mathbf{\mu}_t = \text{Decoder}_\theta(\mathbf{x}_{t-1}, \mathbf{\widehat{z}}^{(q)}_t, \widehat{y}_t)$. Note that
$\mathbf{\widehat{y}}_{1:T} = \{\widehat{y}_1, \cdots, \widehat{y}_T\}$ and $\mathbf{\widehat{z}}^{(q)}_{1:T} = \{\widehat{\mathbf{z}}^{(q)}_1, \cdots, \widehat{\mathbf{z}}^{(q)}_T\}$ are upsampled from $\mathbf{y}_{1:U}$ and $\mathbf{z}^{(q)}_{1:U}$ by duplicating each $y_{u}$ and $z^{(q)}_{u}$ for $l_u$ iterations.

\subsubsection{Prior}
\label{sec:prior}
The prior for $\mathbf{z}^{(q)}_{1:U}$ is defined as
\begin{align}
P_\phi(\mathbf{z}^{(q)}_{1:U} \mid \mathbf{y}_{1:U}) &= \prod_{u=1}^U P(\mathbf{z}^{(q)}_u \mid \mathbf{z}^{(q)}_{u-1}, y_u ), \label{eqn:p_z}
\end{align}
where the probability of observing $\mathbf{z}^{(q)}_u = \mathbf{e}_{l}$ is computed as
\begin{align}
P(\mathbf{z}^{(q)}_u = \mathbf{e}_{l} \mid \mathbf{z}^{(q)}_{u-1}, y_u) &= \frac{\exp(-\|\mathbf{c}_u - \mathbf{e}_{l}\|_2^2)}{\sum_{k=1}^K \exp(-\|\mathbf{c}_u - \mathbf{e}_k\|_2^2)}. 
\end{align}
The activation vector $\mathbf{c}_u$ is computed through a neural network as $\mathbf{c}_u = \mathrm{LatentNet}_\phi(\mathbf{z}^{(q)}_{u-1}, y_u)$. 
Note that the prior calculates the probability of selecting $l$-th codeword $\mathbf{e}_{l}$ for each input token $y_u$, which also decides its duration $l_u = l$. 

Given Eqs.~(\ref{eqn:p_z}) and Eq.~(\ref{eq:indicator}), we compute Eq.~(\ref{eq:elbo_prior})  in the ELBO as
\begin{align}
\label{eq:prior-loss}
&\mathrm{KL}[Q_{\lambda}(\mathbf{z}^{(q)}_{1:U} \mid \mathbf{x}_{1:T}, \mathbf{y}_{1:U}) \| P_\phi(\mathbf{z}^{(q)}_{1:U} \mid \mathbf{y}_{1:U})] \nonumber \\
=& - \log P (\mathbf{z}^{(q)}_{1:U} = (\mathbf{e}_{l_1}, \cdots, \mathbf{e}_{l_U}) \mid \mathbf{y}_{1:U})\\
=& \sum_{u=1}^U \Big\{ \|\mathbf{c}_u - \mathbf{e}_{l_u}\|_2^2 + \log\sum_{k=1}^K \exp(-\|\mathbf{c}_u - \mathbf{e}_k\|_2^2) \Big\}. \label{eq:prior-dictionary-learning}
\end{align}
Since $Q_{\lambda}$ is an indicator function, the value of the KL divergence is equal to the prior models' likelihood on the codeword $\{\mathbf{e}_{l_1}, \cdots, \mathbf{e}_{l_U}\}$ produced by the approximate posterior $Q_{\lambda}$.

\subsubsection{Vector quantization}
\label{sec:vq}
For the vector quantization term in Eq.~(\ref{eq:elbo_vq}), we factorize it as 
\begin{align}
p(\mathbf{{z}}^{(r)}_{1:U} \mid \mathbf{z}^{(q)}_{1:U})=\prod_{u=1}^{U}\mathcal{N}(\mathbf{{z}}^{(r)}_{u}; \mathbf{z}^{(q)}_{u}, \sigma^2\mathbf{I}),
\label{eq:quantization_loss}
\end{align}
and we assume that  $\mathbf{{z}}^{(r)}_{u} = \mathbf{{z}}^{(q)}_{u} + \mathbf{\eta}$, where $\mathbf{\eta} \sim \mathcal{N}(0, \sigma^2\mathbf{I})$. In other words, the quantization loss is Gaussian-distributed. The hyper-parameter $\sigma$ is the same as that in Section~\ref{sec:app_posterior}.

Given Eqs.~(\ref{eq:quantization_loss}) and Eq.~(\ref{eq:f_uni}), we compute Eq.~(\ref{eq:elbo_vq}) in the ELBO as 
\begin{align}
&\mathbb{E}_{Q_{\lambda}(\mathbf{z}^{(q)}_{1:U})}\Big\{\mathrm{KL}\big[{Q_{\psi}(\mathbf{{z}}^{(r)}_{1:U} \mid \mathbf{z}^{(q)}_{1:U})} \| {p(\mathbf{{z}}^{(r)}_{1:U} \mid \mathbf{z}^{(q)}_{1:U}, \mathbf{y}_{1:U})}\big]\Big\}  \nonumber \\
= & \sum_{u=1}^{U} \frac{1}{2\sigma^2} \| \mathbf{d}_u - \mathbf{e}_{l_{u}} \|_2^2 \label{eq:vq_4}.
\end{align}
Note that $Q_\lambda$ is an indicator function, and the KL divergence between two Gaussian distributions has a closed form. We hide $\mathbf{x}_{1:T}, \mathbf{y}_{1:U}$ in $Q_\lambda$ and $Q_\psi$ to avoid notation clutter.

\begin{figure}[!t]
\centering
\includegraphics[trim=10 100 05 260,clip,width=1.0\linewidth]{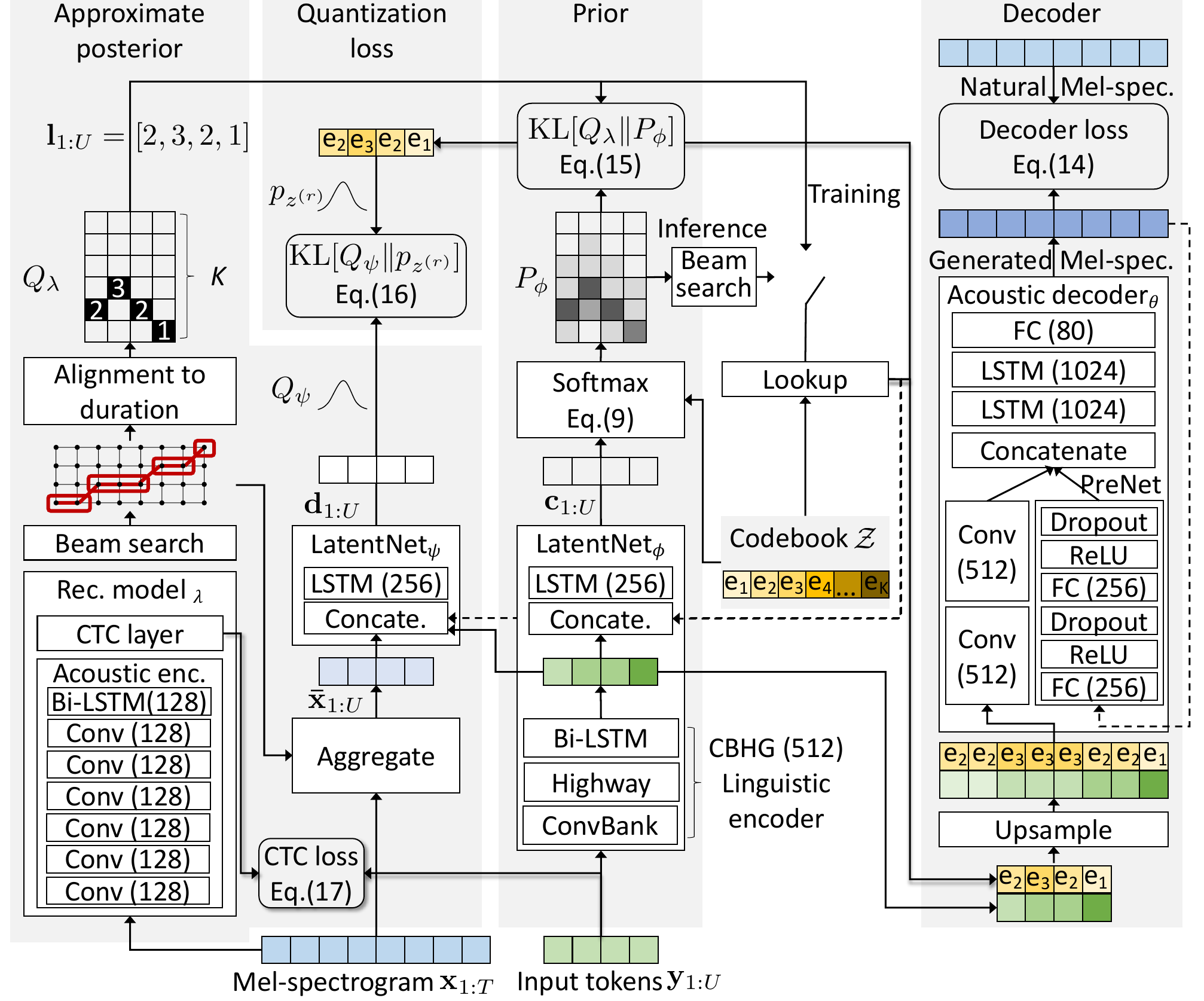}
\vspace{-5mm}
\caption{Proposed TTS system. Dashed line denotes feedback loop. Number in bracket denotes neural layer size. FC denotes a fully connected layer. During inference, only prior and decoder are used.}
\vspace{-5mm}
\label{fig:overview_2}
\end{figure}

\subsection{CTC-based alignment search}
\label{sec:ctc}

In Section~\ref{sec:app_posterior}, we define the approximate posterior $Q_\lambda(\mathbf{z}^{(q)}_{1:U}\mid \mathbf{x}_{1:T}, \mathbf{y}_{1:U})$ as an indicator function. This corresponds to the ideal case where, given $\mathbf{x}_{1:T}$ and $\mathbf{y}_{1:U}$, $Q_\lambda$ can perfectly infer the value of $\mathbf{l}_{1:U}$ and assign probability zero to other values. In practice, we use a CTC-based recognition model $P_\lambda(\mathbf{y}_{1:U} \mid \mathbf{x}_{1:T})$ to produce $\mathbf{l}_{1:U}$. 

This is done by first searching for the optimal alignment $\mathbf{a}^{\ast}_{1:T}$ from the CTC trellis \cite{Graves2006CTC}, i.e.,  
$\mathbf{a^\ast}_{1:T} = \arg\max_{\mathbf{a}_{1:T}}P_\lambda(\mathbf{y}_{1:U}, \mathbf{a}_{1:T} \mid \mathbf{x}_{1:T})$. The alignment variable $a_{t}$ is similar to that in hard-attention in Fig.~\ref{fig:alignment}. We then directly convert $\mathbf{a}^{\ast}_{1:T}$ into $l_{1:U}$ as the example on right side of Fig.~\ref{fig:alignment} shows.

The sampled $l_{1:U}$ must satisfy $T = \sum_u l_{u}$ and $1 \leq l_u \leq K$. This can be achieved by excluding the invalid alignment path from the CTC trellis. This is detailed in the appendix.

\subsection{Training criterion in summary}
\label{sec:training_criteria}

With all the components explained in Sections~\ref{sec:model_parameterization} and \ref{sec:ctc}, we summarize the training criterion of the proposed model as
\begin{align}
&\mathcal{L}(\theta, \psi, \phi, \lambda, \mathcal{Z}) \nonumber \\
= & \mathbb{E}_{Q_{\lambda}(\mathbf{z}^{(q)}_{1:U})}[\log p_\theta(\mathbf{x}_{1:T} \mid \mathbf{z}^{(q)}_{1:U}, \mathbf{y}_{1:U})] \label{eq:loss_1}\\
& - \sum_{u=1}^U \Big\{ \|\mathbf{c}_u - \mathbf{e}_{l_u}\|_2^2 + \log \sum_{k=1}^K \exp({-\|\mathbf{c}_u - \mathbf{e}_k\|_2^2})\Big\} \label{eq:loss_2} \\
&- \sum_{u=1}^{U} \frac{1}{2\sigma^2} \| \mathbf{d}_u -  \mathbf{e}_{l_{u}}\|_2^2  \label{eq:loss_3} \\
&+ \gamma\log \sum_{\forall\mathbf{a}_{1:T}} P_\lambda(\mathbf{y}_{1:U}, \mathbf{a}_{1:T} \mid \mathbf{x}_{1:T}) \label{eq:loss_4}.
\end{align}
Equations ~(\ref{eq:loss_1})--(\ref{eq:loss_3}) correspond to the ELBO in Section~\ref{sec:model_def}, and the last line is for the CTC model. 
The vectors $\mathbf{d}_u$ and $\mathbf{c}_u$ are computed as $\mathbf{d}_u = \mathrm{LatentNet}_\psi(\mathbf{z}^{(q)}_{u-1}, \mathbf{\bar{x}}_u, y_u)$ and $\mathbf{c}_u = \mathrm{LatentNet}_\phi(\mathbf{z}^{(q)}_{u-1}, y_u)$.

The decoder parameter $\theta$ is trained with Eq.~(\ref{eq:loss_1}); the $\mathrm{LatentNet}_\phi$ in prior is trained with Eq.~(\ref{eq:loss_2}); and the $\mathrm{LatentNet}_\psi$ in approximate posterior is trained with Eq.~(\ref{eq:loss_3}). The codebook $\mathcal{Z}=\{\mathbf{e}_1, \cdots, \mathbf{e}_K\}$ is trained with Eqs.~(\ref{eq:loss_2}--\ref{eq:loss_3}). All the components are jointly trained. During inference, only the prior and decoder network are used. 

\subsection{Implementation}
\label{sec:tts}
The implemented TTS system is shown in Fig.~\ref{fig:overview_2}. 
The acoustic encoder in the CTC model consists of convolution layers followed by a bidirectional LSTM, which is similar to that in \cite{DBLP:conf/interspeech/KlimkovRRD19}. The $\mathrm{LatentNet}_\phi$ in the prior consists of an CBHG-based linguistic encoder \cite{Wang2017} and an additional LSTM layer. 
$\mathrm{LatentNet}_\psi$ has only one LSTM layer. The decoder is similar to that in \cite{Wang2017}, which consists of pre-net bottleneck layers,  convolutional layers, LSTM, and linear output layers. 
The decoder receives input from the linguistic encoder.  
A more detailed explanation is provided in the appendix.

\vspace{-5pt}
\section{Experiment}
\label{sec:experiments}

\subsection{Experimental conditions}

In the experiment, we treated the duration in three frames as one unit. This grouping allowed us to reduce the codebook size $K$ to 13. Accordingly, the value of the latent duration can be $l_u=k$, where $k \in \{3, 6, \cdots, 39\}$. Each codebook vector $e_k$ has 32 dimensions.  

We configured $\sigma_d = 3.0$ in Eq.~(\ref{eq:decoder}), and $\sigma = 0.4$ in Eq.~(\ref{eq:vq_4}). Square distance in Eqs.~(\ref{eq:prior-dictionary-learning}) and ~(\ref{eq:vq_4}) was decomposed into $f(a, b) = \alpha \|a - \mathrm{sg}[b]\|_2^2 + \beta \|\mathrm{sg}[a] - b\|_2^2$, where $\mathrm{sg}[\cdot]$ is a stop gradient operator. We configured $\alpha = 1.0, \beta = 0.0$ for Eq.~(\ref{eq:prior-dictionary-learning}), and $\alpha = 2.0, \beta = 1.0$ for Eq.~(\ref{eq:vq_4}). We set $\gamma = 0.5$ for CTC loss in Eq.~(\ref{eq:loss_4}). We used the Adam optimizer \cite{DBLP:journals/corr/KingmaB14} with a fixed learning rate of $5 \times 10^{-5}$ to optimize the objective. Durations were sampled with a beam width of 3 during training and of 10 during inference.

We used LJSpeech\footnote{https://keithito.com/LJ-Speech-Dataset/}, an English speech corpus containing about 24 hours of recordings from a female speaker. We used 12,600 utterances for training, 250 for validation, and 250 for testing. 

We built three versions of the proposed system that use character or phoneme as the input tokens $\mathbf{y}_{1:U}$:
\begin{itemize}
    \item \texttt{CC} uses characters as input tokens;
    \item \texttt{PP} uses phonemes as input tokens; and
    \item \texttt{CP} uses characters as the input of the linguistic encoder and character-aligned phonemes as the CTC target.
\end{itemize}
The phoneme labels are obtained with a G2P module\footnote{https://github.com/Kyubyong/g2p}. All systems use 80-dimensional Mel-spectrogram. 
Waveform were generated with WaveNet \cite{oord2016wavenet}\footnote{For fair comparison with TTS systems, we used a public WaveNet called ljspeech.wavenet.mol.v1 provided by ESPNet.}.

We conducted a listening test to evaluate the performance of our proposed TTS system. We included nine systems in the listening test: natural samples (\texttt{Natural}), analysis-by-synthesis (\texttt{ABS}), Transformer-TTS \cite{DBLP:conf/aaai/Li0LZL19}, FastSpeech \cite{Ren2019}, two Tacotron2 models \cite{Shen2017} as reference systems, and the three proposed systems\footnote{Audio samples are available at \url{https://nii-yamagishilab.github.io/sample-tts-latent-duration/}}. The reference systems are public models built by the ESPNet-TTS team \cite{DBLP:conf/icassp/HayashiYIY0TTZT20}: \texttt{transformer.v3}, \texttt{fastspeech.v3}, \texttt{tacotron2.v2}, and \texttt{tacotron2.v3}.
We selected these reference systems in order to compare our proposed system with TTS systems that use different duration modeling approaches; specifically, FastSpeech uses an external duration model, Tacotron-based systems use soft-attention, and Transformer-TTS uses soft-attention with positional encoding.

We recruited 200 Japanese listeners through crowdsourcing. In each listening set, one listener was asked to evaluate the quality of 36 audio samples using a five-grade mean-opinion-score (MOS) scale. We collected 17,856 evaluation scores in total.

\vspace{-5pt}
\subsection{Experimental results}

\begin{figure}[!t]
\centering
\includegraphics[width=1.0\linewidth]{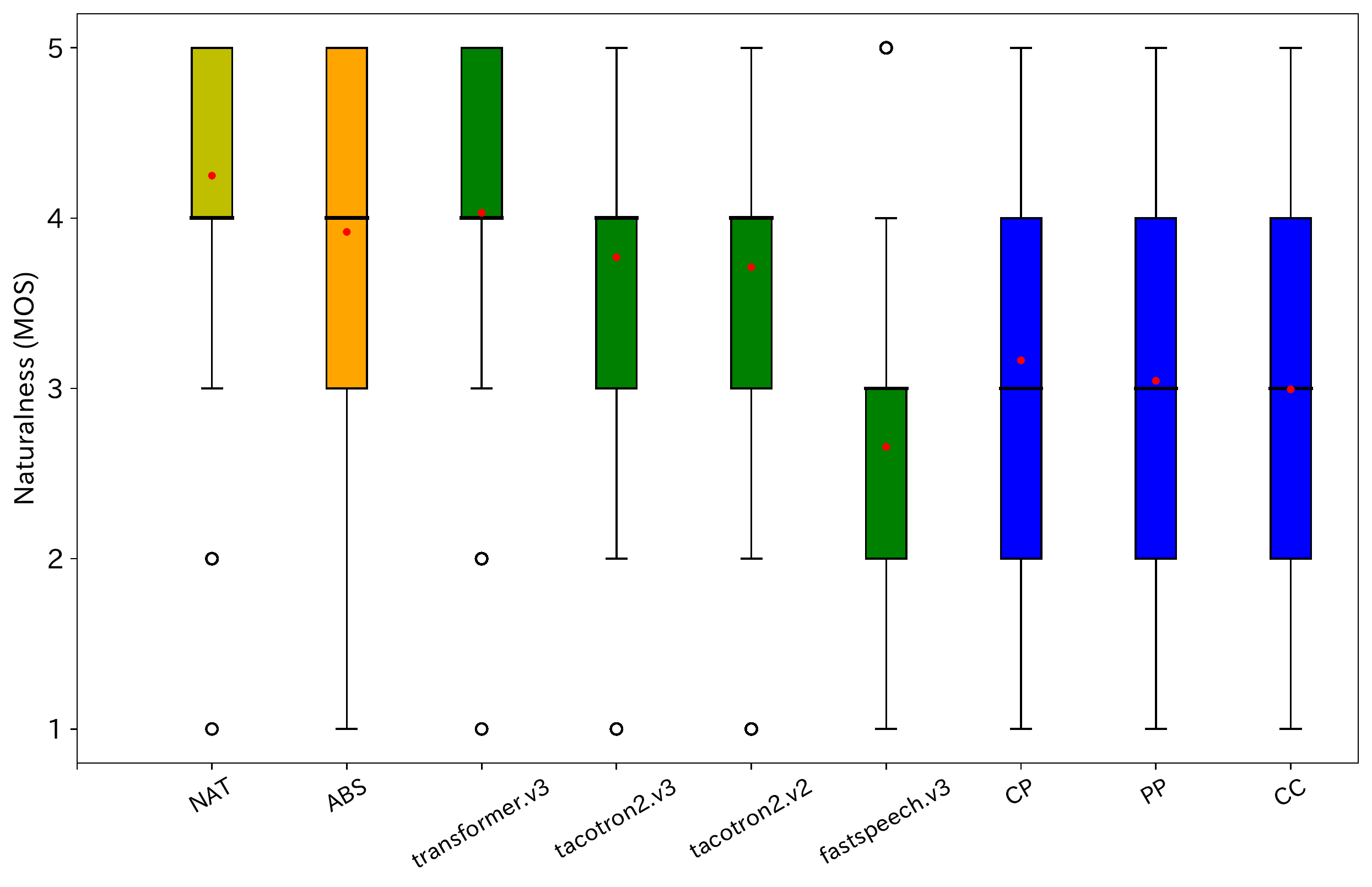}
\caption{Results of listening test. Red dots denote mean MOS values.}
\label{fig:listening-test}
\end{figure}

Figure \ref{fig:listening-test} shows the results of the listening test. Our proposed systems got $2.99 \pm 0.04$ for the character-based system (\texttt{CC}), $3.04 \pm 0.04$  for phoneme-based system (\texttt{PP}), and $3.16 \pm 0.04$ for the character and phoneme combination system (\texttt{CP}). For reference systems, \texttt{transformer.v3}, \texttt{tacotron2.v3}, \texttt{tacotron2.v2}, and \texttt{fastspeech.v3} got $4.03 \pm 0.03$, $3.77 \pm 0.03$, $3.71 \pm 0.03$, and $2.66 \pm 0.04$, respectively.

Overall, our proposed systems were rated between the soft-attention-based systems (Transformer-TTS and Tacotron2) and FastSpeech that uses an external duration model instead of soft-attention. Among the proposed systems, the character and phoneme combination system (\texttt{CP}) obtained the best score. We found that phoneme labels estimated by the G2P module contained errors, which resulted in mispronunciations of the synthesized speech in the \texttt{PP} condition. 

Although our proposed systems were rated worse than the soft-attention-based systems, we argue that this is not surprising, as there was a strong assumption in our model. Specifically, the current framework automatically estimates the duration of given input symbols, but it does not insert any short pauses between the symbols. It does not determine the duration of the short pauses, either. In other words, we set a constraint that the sum of duration of each input character is equal to speech duration, that is, $T = \sum_u l_{u}$, and this constraint does not consider the duration of any short pauses, which may be inserted into any phrase boundary. Obviously this makes synthesized speech unnatural perceptually, since the synthetic speech of the proposed system has neither short pauses nor phrase breaks. Our model needs to have a more appropriate constraint where we exclude the total duration of short pauses and an additional mechanism to insert short pauses at appropriate phrase boundaries. This is our next step.      



\vspace{-10pt}
\section{Conclusion}
\label{sec:conclusion}
We proposed a sequence-to-sequence TTS system that treats duration as a discrete latent variable. 
During training, we can conceptually interpret the approximate posterior network as a force aligner and the prior network as a text-to-duration model. However, all the components are jointly trained from scratch by maximizing a theoretically derived variational lower bound. 
For generation, the prior extracts linguistic features and predicts the latent duration, and the decoder generates the acoustic features from the input text. We experimentally compared three versions of our proposed TTS method with other sequence-to-sequence TTS systems, including FastSpeech, Tacotron2, and Transformer-TTS, and found that our systems had better naturalness than FastSpeech but were rated lower than Transformer-TTS and Tacotron2. We presume that the inferior naturalness was caused by the lack of appropriate modeling of short pauses. One possible solution to improve our systems is to automatically determine short pauses during training and insert them at inference time. This will be the focus of our future work.


\vfill\pagebreak

\bibliographystyle{IEEEbib}
\bibliography{sample}

\clearpage
\appendix
\onecolumn
\def\x{{\mathbf x}}
\def\L{{\cal L}}

\def\ci{\perp\!\!\!\perp}






\ninept


\section{Variational lower bound for proposed model}
\subsection{ELBO in detail}
To recap, we use $\mathbf{x}_{1:T}$ and $\mathbf{y}_{1:U}$ to denote the output acoustic feature and input linguistic (e.g., phoneme and character) sequences, respectively. We use $\mathbf{z}^{(q)}_{u}$ to denote the quantized latent and $\mathbf{z}^{(r)}_{u}$ to denote the un-quantized latent vectors for the $u$-th input token, $\forall{u}\in\{1,\cdots,U\}$. We further assume $\mathbf{z}^{(r)}_{u}\in\mathbb{R}^{D}$, where $D$ is the dimension of the latent vector. For the quantized latent variable, we let $\mathbf{z}^{(q)}_{u}\in\mathcal{Z}=\{\mathbf{e}_1, \cdots, \mathbf{e}_K\}$, where $\mathcal{Z}$ is the codebook with $K$ code words, and $\mathbf{e}_k \in \mathbb{R}^{D}, \forall{k}\in\{1, \cdots, K\}$.

We treat $\mathbf{z}^{(q)}_{1:U}=(\mathbf{z}^{(q)}_{1}, \cdots, \mathbf{z}^{(q)}_{U})$ and $\mathbf{z}^{(r)}_{1:U}=(\mathbf{z}^{(r)}_{1}, \cdots, \mathbf{z}^{(r)}_{U})$ as latent variables. With Jensen's inequality, we have
\begin{align}
&\log p(\mathbf{x}_{1:T} \mid \mathbf{y}_{1:U}) \\
= & \log\sum_{\forall \mathbf{z}^{(q)}_{1:U}}\int_{\mathbf{{z}}^{(r)}_{1:U}} p(\mathbf{x}_{1:T}, \mathbf{z}^{(q)}_{1:U}, \mathbf{{z}}^{(r)}_{1:U} \mid \mathbf{y}_{1:U}) d\mathbf{{z}}^{(r)}_{1:U} \label{app_eq:model_exact}\\
= & \log\sum_{\forall \mathbf{z}^{(q)}_{1:U}} \int_{\mathbf{z}^{(r)}_{1:U}} \underbrace{p_{\theta}(\mathbf{x}_{1:T} \mid \mathbf{z}^{(q)}_{1:U}, \mathbf{y}_{1:U})}_{(\mathbf{x}_{1:T}\ci\mathbf{z}^{(r)}_{1:U}) | \mathbf{z}^{(q)}_{1:U} } \, p(\mathbf{{z}}^{(r)}_{1:U} \mid \mathbf{{z}}^{(q)}_{1:U}, \mathbf{y}_{1:U}) P_\phi(\mathbf{z}^{(q)}_{1:U} \mid \mathbf{y}_{1:U}) d\mathbf{{z}}^{(r)}_{1:U} \label{app_eq:model_assumption}\\
\ge &\sum_{\forall \mathbf{z}^{(q)}_{1:U}} \int_{\mathbf{z}^{(r)}_{1:U}} \underbrace{Q(\mathbf{z}^{(q)}_{1:U}, \mathbf{{z}}^{(r)}_{1:U} \mid \mathbf{x}_{1:T}, \mathbf{y}_{1:U})}_{\text{approximate posterior}}\log \frac{ p_{\theta}(\mathbf{x}_{1:T} \mid \mathbf{z}^{(q)}_{1:U}, \mathbf{y}_{1:U}) \, p(\mathbf{{z}}^{(r)}_{1:U} \mid \mathbf{{z}}^{(q)}_{1:U}, \mathbf{y}_{1:U}) P_\phi(\mathbf{z}^{(q)}_{1:U} \mid \mathbf{y}_{1:U}) }{\underbrace{Q(\mathbf{z}^{(q)}_{1:U}, \mathbf{{z}}^{(r)}_{1:U} \mid \mathbf{x}_{1:T}, \mathbf{y}_{1:U})}_{\text{approximate posterior, factorize it to } Q(z^{(q)})Q(z^{(r)}|z^{(q)})}} d\mathbf{{z}}^{(r)}_{1:U} \label{app_eq:elbo} \\
=& \sum_{\forall \mathbf{z}^{(q)}_{1:U}} \int_{\mathbf{z}^{(r)}_{1:U}} Q(\mathbf{z}^{(q)}_{1:U}, \mathbf{{z}}^{(r)}_{1:U} \mid \mathbf{x}_{1:T}, \mathbf{y}_{1:U}) \underbrace{\log \frac{ p_{\theta}(\mathbf{x}_{1:T} \mid \mathbf{z}^{(q)}_{1:U}, \mathbf{y}_{1:U}) \, p(\mathbf{{z}}^{(r)}_{1:U} \mid \mathbf{{z}}^{(q)}_{1:U}, \mathbf{y}_{1:U}) P_\phi(\mathbf{z}^{(q)}_{1:U} \mid \mathbf{y}_{1:U}) }{
Q_{\psi}(\mathbf{{z}}^{(r)}_{1:U} \mid \mathbf{z}^{(q)}_{1:U},  \mathbf{x}_{1:T}, \mathbf{y}_{1:U}) Q_{\lambda}(\mathbf{z}^{(q)}_{1:U} \mid \mathbf{x}_{1:T}, \mathbf{y}_{1:U}) }}_{\text{decompose this term}} d\mathbf{{z}}^{(r)}_{1:U} \\
=& \sum_{\forall \mathbf{z}^{(q)}_{1:U}} \underbrace{\int_{\mathbf{z}^{(r)}_{1:U}} Q(\mathbf{z}^{(q)}_{1:U}, \mathbf{{z}}^{(r)}_{1:U} \mid \mathbf{x}_{1:T}, \mathbf{y}_{1:U})}_{\text{integrate out}\, \mathbf{{z}}^{(r)}_{1:U}} \log p_{\theta}(\mathbf{x}_{1:T} \mid \mathbf{z}^{(q)}_{1:U}, \mathbf{y}_{1:U}) d\mathbf{{z}}^{(r)}_{1:U}  \\
&+\sum_{\forall \mathbf{z}^{(q)}_{1:U}} \underbrace{\int_{\mathbf{z}^{(r)}_{1:U}} Q(\mathbf{z}^{(q)}_{1:U}, \mathbf{{z}}^{(r)}_{1:U} \mid \mathbf{x}_{1:T}, \mathbf{y}_{1:U})}_{\text{integrate out}\, \mathbf{{z}}^{(r)}_{1:U}}\log \frac{ P_\phi(\mathbf{z}^{(q)}_{1:U} \mid \mathbf{y}_{1:U}) }{
Q_{\lambda}(\mathbf{z}^{(q)}_{1:U} \mid \mathbf{x}_{1:T}, \mathbf{y}_{1:U})} d\mathbf{{z}}^{(r)}_{1:U} \\
&+\sum_{\forall \mathbf{z}^{(q)}_{1:U}} \int_{\mathbf{z}^{(r)}_{1:U}} \underbrace{Q(\mathbf{z}^{(q)}_{1:U}, \mathbf{{z}}^{(r)}_{1:U} \mid \mathbf{x}_{1:T}, \mathbf{y}_{1:U})}_{Q(\mathbf{z}^{(q)}_{1:U})Q(\mathbf{z}^{(r)}_{1:U} \mid\mathbf{z}^{(q)}_{1:U})}\log \frac{  p(\mathbf{{z}}^{(r)}_{1:U} \mid \mathbf{{z}}^{(q)}_{1:U}, \mathbf{y}_{1:U}) }{
Q_{\psi}(\mathbf{{z}}^{(r)}_{1:U} \mid \mathbf{z}^{(q)}_{1:U},  \mathbf{x}_{1:T}, \mathbf{y}_{1:U})} d\mathbf{{z}}^{(r)}_{1:U} \\
=& \mathbb{E}_{Q_{\lambda}(\mathbf{z}^{(q)}_{1:U} \mid \mathbf{x}_{1:T}, \mathbf{y}_{1:U})}[{\log p_\theta(\mathbf{x}_{1:T} \mid \mathbf{z}^{(q)}_{1:U}, \mathbf{y}_{1:U})}] \\
&+\sum_{\forall \mathbf{z}^{(q)}_{1:U}} Q_{\lambda}(\mathbf{z}^{(q)}_{1:U} \mid \mathbf{x}_{1:T}, \mathbf{y}_{1:U}) \log \frac{ P_\phi(\mathbf{z}^{(q)}_{1:U} \mid \mathbf{y}_{1:U}) }{
Q_{\lambda}(\mathbf{z}^{(q)}_{1:U} \mid \mathbf{x}_{1:T}, \mathbf{y}_{1:U})} d\mathbf{{z}}^{(r)}_{1:U} \\
&+\sum_{\forall \mathbf{z}^{(q)}_{1:U}} Q_{\lambda}(\mathbf{z}^{(q)}_{1:U} \mid \mathbf{x}_{1:T}, \mathbf{y}_{1:U})  \int_{\mathbf{z}^{(r)}_{1:U}} Q_{\psi}( \mathbf{{z}}^{(r)}_{1:U} \mid \mathbf{z}^{(q)}_{1:U}, \mathbf{x}_{1:T}, \mathbf{y}_{1:U})\log \frac{  p(\mathbf{{z}}^{(r)}_{1:U} \mid \mathbf{{z}}^{(q)}_{1:U}, \mathbf{y}_{1:U}) }{
Q_{\psi}(\mathbf{{z}}^{(r)}_{1:U} \mid \mathbf{z}^{(q)}_{1:U},  \mathbf{x}_{1:T}, \mathbf{y}_{1:U})} d\mathbf{{z}}^{(r)}_{1:U} \\
= &\mathbb{E}_{Q_{\lambda}(\mathbf{z}^{(q)}_{1:U} \mid \mathbf{x}_{1:T}, \mathbf{y}_{1:U})}[\underbrace{\log p_\theta(\mathbf{x}_{1:T} \mid \mathbf{z}^{(q)}_{1:U}, \mathbf{y}_{1:U})}_{\text{Decoder}}] \label{app_eq:elbo_decoder}\\
~ &-\mathrm{KL}[{Q_{\lambda}(\mathbf{z}^{(q)}_{1:U} \mid \mathbf{x}_{1:T}, \mathbf{y}_{1:U})} \| \underbrace{P_\phi(\mathbf{z}^{(q)}_{1:U} \mid \mathbf{y}_{1:U})}_{\text{Prior}}] \label{app_eq:elbo_prior} \\ 
~ &- \mathbb{E}_{Q_{\lambda}(\mathbf{z}^{(q)}_{1:U} \mid \mathbf{x}_{1:T}, \mathbf{y}_{1:U})} \Big\{\mathrm{KL}[{Q_{\psi}(\mathbf{{z}}^{(r)}_{1:U} \mid \mathbf{z}^{(q)}_{1:U}, \mathbf{x}_{1:T}, \mathbf{y}_{1:U})} \| \underbrace{p(\mathbf{{z}}^{(r)}_{1:U} \mid \mathbf{z}^{(q)}_{1:U}, \mathbf{y}_{1:U})}_{\text{Vector quantization}}] \Big\} \label{app_eq:elbo_vq} \\
= & \text{ELBO}(\mathbf{x}_{1:T}, \mathbf{y}_{1:U})
\end{align}
Equations~(\ref{app_eq:elbo_decoder})--(\ref{app_eq:elbo_vq}) denote the evidence lower bound (ELBO).
In Eq.~(\ref{app_eq:model_assumption}), we assume $(\mathbf{x}_{1:T}\ci\mathbf{z}^{(r)}_{1:U}) | \mathbf{z}^{(q)}_{1:U} $, which means that $\mathbf{x}_{1:T}$ is conditionally independent of $\mathbf{z}^{(r)}_{1:U}$, given $\mathbf{z}^{(q)}_{1:U}$ and $\mathbf{y}_{1:U}$. We use the chain rule to factorize the distribution
$p(\mathbf{x}_{1:T}, \mathbf{z}^{(q)}_{1:U}, \mathbf{{z}}^{(r)}_{1:U} \mid \mathbf{y}_{1:U}) = p_{\theta}(\mathbf{x}_{1:T} \mid \mathbf{z}^{(q)}_{1:U}, \mathbf{y}_{1:U})\, p(\mathbf{{z}}^{(r)}_{1:U} \mid \mathbf{{z}}^{(q)}_{1:U}, \mathbf{y}_{1:U}) P_\phi(\mathbf{z}^{(q)}_{1:U} \mid \mathbf{y}_{1:U})$. The definition of each part of the ELBO is explained in the following subsections.

\subsection{Approximate posterior}
We define $Q(\mathbf{z}^{(q)}_{1:U}, \mathbf{z}^{(r)}_{1:U} \mid \mathbf{y}_{1:U}, \mathbf{x}_{1:T})$ as 
\begin{equation}
Q(\mathbf{z}^{(q)}_{1:U}, \mathbf{z}^{(r)}_{1:U} \mid \mathbf{y}_{1:U}, \mathbf{x}_{1:T}) = Q_{\psi}(\mathbf{z}^{(r)}_{1:U} \mid \mathbf{z}^{(q)}_{1:U}, \mathbf{y}_{1:U}, \mathbf{x}_{1:T}) Q_{\lambda}(\mathbf{z}^{(q)}_{1:U} \mid \mathbf{y}_{1:U}, \mathbf{x}_{1:T}),
\label{app_eq:pos1}
\end{equation}
where  
\begin{align}
Q_{\lambda}(\mathbf{z}^{(q)}_{1:U} \mid \mathbf{y}_{1:U}, \mathbf{x}_{1:T}) = \prod_{u=1}^U P(\mathbf{z}^{(q)}_u \mid \mathbf{y}_{1:U}, \mathbf{x}_{1:T}) =  \prod_{u=1}^U\mathcal{I}(\mathbf{z}^{(q)}_u = \mathbf{e}_{l_u}), \label{app_eq:indicator}
\end{align}
and
\begin{align}Q_{\psi}(\mathbf{z}^{(r)}_{1:U} \mid \mathbf{z}^{(q)}_{1:U}, \mathbf{y}_{1:U}, \mathbf{x}_{1:T}) =& \prod_{u=1}^U p(\mathbf{z}^{(r)}_u \mid  \mathbf{z}^{(q)}_{u-1}, \mathbf{y}_{1:U}, \mathbf{x}_{1:T}) = \prod_{u=1}^U \mathcal{F}(\mathbf{z}^{(r)}_{u} -  \mathbf{d}_u), \label{app_eq:f_uni}\\ 
\mathbf{d}_{u} &=\mathrm{LatentPredictor}_\psi(\mathbf{z}^{(q)}_{u-1}, \mathbf{\bar{x}}_u, y_u). \label{app_eq:q_d}
\end{align}
In the above equations, $l_u$ is the codeword index for the $u$-th input token, $\mathcal{I}(\cdot)$ is an indicator function, and $\mathcal{F}(\cdot)$ is any fixed unimodal probability density function centered on the origin, such as the isotropic Gaussian $\mathcal{F}(\mathbf{x}) = \mathcal{N}(\mathbf{x}; 0, \mathbf{I})$. The LatentPredictor is a neural network with a parameter set $\psi$.

Note that $\mathbf{\bar{x}}_u$ is from an aggregated acoustic features sequence $\mathbf{\bar{x}}_{1:U} = (\mathbf{\bar{x}}_1, \cdots, \mathbf{\bar{x}}_U)$, which has the same length $U$ as the linguistic feature input $\mathbf{\bar{y}}_{1:U}$. The aggregation is conducted on the basis of the value of codeword indices. Suppose the value of $\mathbf{z}^{(q)}_{1:U}$ is $(\mathbf{e}_{l_1}, \cdots,\mathbf{e}_{l_U})$. Then the aggregation can be written as
\begin{align}
\mathbf{\bar{x}}_{1:U} = \mathrm{Aggregate}(\mathbf{x}_{1:T}, \mathbf{l}_{1:U}) = \left\{\mathbf{\bar{x}}_u \middle| \mathbf{\bar{x}}_u = \sum_{t \in \mathcal{T}(l_{1:u})}\mathbf{x}_t/l_u \right\} \label{app_eq:aggregate},
\end{align}
where $\mathbf{l}_{1:U}=(l_1, \cdots, l_U)$ is the index sequence. The function $\mathcal{T}(l_{1:u})=\{t | t\in[\mathcal{T}(l_{1:(u-1)})+1, \mathcal{T}(l_{1:(u-1)}) + l_u]\}$ returns the output time steps that belong to the $u$-th input token, given the initial condition $\mathcal{T}(l_{1:0})=0$. For example, if we know $l_1 = 2$ and $l_2 = 3$, $\mathcal{T}(l_{1:1})=\{1, 2\}$ and $\mathcal{T}(l_{1:2})=\{3, 4, 5\}$.

\subsection{Decoder}
Just to repeat the definition of the decoder, given $\mathbf{y}_{1:U}$, $\mathbf{l}_{1:U}=(l_1, \cdots, l_U)$ and $\mathbf{z}^{(q)}_{1:U}=(\mathbf{z}^{(q)}_{1}=\mathbf{e}_{l_1}, \cdots, \mathbf{z}^{(q)}_{U}=\mathbf{e}_{l_U})$, the PDF of  $\mathbf{x}_{1:T}$ is defined as
\begin{align} 
p_\theta(\mathbf{x}_{1:T} \mid \mathbf{z}^{(q)}_{1:U}, \mathbf{y}_{1:U}) = \prod_{t=1}^T p(\mathbf{x}_t\mid \mathbf{x}_{t-1}, \mathbf{\widehat{z}}^{(q)}_t, \widehat{y}_t) = \prod_{t=1}^T \mathcal{N}(\mathbf{x}_t; \mathbf{\mu}_t, \sigma_d^2\mathbf{I}), \label{app_eq:decoder}
\end{align}
where $\mathbf{\mu}_t = \text{Decoder}_\theta(\mathbf{x}_{t-1}, \mathbf{\widehat{z}}^{(q)}_t, \widehat{y}_t)$. Note that
$\mathbf{\widehat{y}}_{1:T} = \{\widehat{y}_1, \cdots, \widehat{y}_T\}$ and $\mathbf{\widehat{z}}^{(q)}_{1:T} = \{\widehat{\mathbf{z}}^{(q)}_1, \cdots, \widehat{\mathbf{z}}^{(q)}_T\}$ are upsampled from $\mathbf{y}_{1:U}$ and $\mathbf{z}^{(q)}_{1:U}$ by duplicating each $\mathbf{y}_{u}$ and $\mathbf{y}_{u}$ for $l_u$ iterations. 

\subsection{Prior}
We assume the prior probability of observing the $l$-th codeword $\mathbf{e}_{l}$ to be
\begin{align}
P_\phi(\mathbf{z}^{(q)}_{1:U} = \mathbf{e}_{l} \mid \mathbf{y}_{1:U}) &= \prod_{u=1}^U P(\mathbf{z}^{(q)}_u =\mathbf{e}_{l} \mid \mathbf{z}^{(q)}_{u-1}, y_u ) = \prod_{u=1}^U \frac{\exp(-\|\mathbf{c}_u - \mathbf{e}_{l}\|_2^2)}{\sum_{k=1}^K \exp(-\|\mathbf{c}_u - \mathbf{e}_k\|_2^2)}, \label{app_eqn:p_z}  \\
\mathbf{c}_u &= \mathrm{LatentPredictor}_\phi(\mathbf{z}^{(q)}_{u-1}, y_u),
\end{align}
where $\mathrm{LatentPredictor}_\phi(\mathbf{z}^{(q)}_{u-1}, y_u) $ is a neural network with parameter $\phi$.

With Eqs.~(\ref{app_eqn:p_z}) and Eq.~(\ref{app_eq:indicator}), the KL divergence in Eq.~(\ref{app_eq:elbo_prior}) can be computed as
\begin{align}
\label{app_eq:prior-loss}
\mathrm{KL}[Q_{\lambda}(\mathbf{z}^{(q)}_{1:U} \mid \mathbf{x}_{1:T}, \mathbf{y}_{1:U}) \| P_\phi(\mathbf{z}^{(q)}_{1:U} \mid \mathbf{y}_{1:U})] =& \mathbb{E}_{Q_\lambda(\mathbf{z}^{(q)}_{1:U} \mid\mathbf{x}_{1:T}, \mathbf{y}_{1:U})}[\log \frac{Q_{\lambda}(\mathbf{z}^{(q)}_{1:U} \mid \mathbf{x}_{1:T}, \mathbf{y}_{1:U})}{P_{\phi}(\mathbf{z}^{(q)}_{1:U} \mid \mathbf{y}_{1:U})}], \\
=& - \mathbb{E}_{Q_\lambda(\mathbf{z}^{(q)}_{1:U} \mid\mathbf{x}_{1:T}, \mathbf{y}_{1:U})}[\log P_{\phi}(\mathbf{z}^{(q)}_{1:U} \mid \mathbf{y}_{1:U})] - \underbrace{H(Q_\lambda(\mathbf{z}^{(q)}_{1:U} \mid\mathbf{x}_{1:T}, \mathbf{y}_{1:U}))}_{=0\text{ for indicator function}},\\
\label{app_eqn:kld}
=& - \log P_{\phi}(\mathbf{z}^{(q)}_{1:U} = (\mathbf{e}_{l_1}, \cdots, \mathbf{e}_{l_U}) \mid \mathbf{y}_{1:U}),\\
=& \sum_{u=1}^U \|\mathbf{c}_u - \mathbf{e}_{l_u}\|_2^2 + \log\sum_{k=1}^K \exp(-\|\mathbf{c}_u - \mathbf{e}_k\|_2^2). \label{app_eq:prior-dictionary-learning}
\end{align}
Note that the $H(\cdot)$ term calculates the entropy of the indicator function, which is equal to 0.

\subsection{Vector quantization}
\label{app_seq:app_vq}
In the final term in Eq.~(\ref{app_eq:elbo_vq}), let us define 
\begin{align}
p(\mathbf{{z}}^{(r)}_{1:U} \mid \mathbf{z}^{(q)}_{1:U}, \mathbf{y}_{1:U}) = \prod_{u=1}^{U}p(\mathbf{{z}}^{(r)}_{u} \mid \mathbf{z}^{(q)}_{u})=\prod_{u=1}^{U}\mathcal{N}(\mathbf{{z}}^{(r)}_{u}; \mathbf{z}^{(q)}_{u}, \sigma^2\mathbf{I}), 
\label{app_eq:quantization_loss}
\end{align}
where $\mathbf{I}$ is an identity matrix and $\sigma^2$ is a fixed parameter. For the first equality, we assume that $\mathbf{{z}}^{(r)}_{u}$ is conditionally independent from $\mathbf{y}_{1:U}$, $\mathbf{z}^{(q)}_{v}$, and $\mathbf{z}^{(r)}_{v}, {v}{\neq{u}}$. This second equality assumes that $\mathbf{{z}}^{(r)}_{u} = \mathbf{{z}}^{(q)}_{u} + \mathbf{\eta}$, where $\mathbf{\eta} \sim \mathcal{N}(0, \sigma^2\mathbf{I})$. In this case, we assume the quantization loss is from a Gaussian distribution.

With Eqs.~(\ref{app_eq:f_uni}) and Eq.~(\ref{app_eq:quantization_loss}), we can compute Eq.~(\ref{app_eq:elbo_vq}) in ELBO as
\begin{align}
&\mathbb{E}_{\underbrace{Q_{\lambda}(\mathbf{z}^{(q)}_{1:U} \mid \mathbf{x}_{1:T}, \mathbf{y}_{1:U})}_{\text{indicator function}}}\Big\{\mathrm{KL}\big[{Q_{\psi}(\mathbf{{z}}^{(r)}_{1:U} \mid \mathbf{z}^{(q)}_{1:U}, \mathbf{x}_{1:T}, \mathbf{y}_{1:U})} \| {p(\mathbf{{z}}^{(r)}_{1:U} \mid \mathbf{z}^{(q)}_{1:U}, \mathbf{y}_{1:U})}\big]\Big\},   \\
= &\mathrm{KL}\big[{Q_{\psi}(\mathbf{{z}}^{(r)}_{1:U} \mid \mathbf{z}^{(q)}_{1:U} = \mathbf{e}_{1:U}, \mathbf{x}_{1:T}, \mathbf{y}_{1:U})} \| {p(\mathbf{{z}}^{(r)}_{1:U} \mid \mathbf{z}^{(q)}_{1:U}=\mathbf{e}_{1:U}, \mathbf{y}_{1:U})}\big] \label{app_eq:vq_1},\\
= & \sum_{u=1}^{U} \mathrm{KL}\big[ p(\mathbf{z}^{(r)}_u \mid \mathbf{z}^{(q)}_{u-1}=\mathbf{e}_{l_{u-1}}, \mathbf{y}_{1:U}, \mathbf{x}_{1:T}) \|  \mathcal{N}(\mathbf{{z}}^{(r)}_{u}; \mathbf{z}^{(q)}_{u}=\mathbf{e}_{l_{u}}, \sigma^2\mathbf{I}) \big] \label{app_eq:vq_2}.
\end{align}

In Eq.~(\ref{app_eq:f_uni}), we defined that $p(\mathbf{z}^{(r)}_u \mid \mathbf{z}^{(q)}_{u-1}, \mathbf{y}_{1:U}, \mathbf{x}_{1:T}) = \mathcal{F}(\mathbf{z}^{(r)}_{u} -  \mathbf{d}_u)$, where $\mathcal{F}(\cdot)$ can be a unimodal distribution and $\mathbf{d}_u = \mathrm{LatentPredictor}_\psi(\mathbf{z}^{(q)}_{u-1}, \mathbf{\bar{x}}_u, y_u)$. 
Now let us define $\mathcal{F}(\mathbf{x})=\mathcal{N}(\mathbf{x}; 0, \sigma^2\mathbf{I})$, whose covariance matrix is the same as that for $\mathcal{N}(\mathbf{{z}}^{(r)}_{u}; \mathbf{z}^{(q)}_{u}, \sigma^2\mathbf{I})$. Accordingly, we have $p(\mathbf{z}^{(r)}_u \mid \mathbf{z}^{(q)}_{u-1}, \mathbf{y}_{1:U}, \mathbf{x}_{1:T}) =\mathcal{N}(\mathbf{{z}}^{(r)}_{u}; \mathbf{d}_u, \sigma^2\mathbf{I}) $. The KL divergence between the two multivariate Gaussians can be analytically computed as
\begin{align}
\mathrm{KL}\big[  \mathcal{N}(\mathbf{{z}}^{(r)}_{u}; \mathbf{d}_u, \sigma^2\mathbf{I}) \|  \mathcal{N}(\mathbf{{z}}^{(r)}_{u}; \mathbf{z}^{(q)}_{u}=\mathbf{e}_{l_{u}}, \sigma^2\mathbf{I}) \big] = \frac{1}{2\sigma^2} \| \mathbf{d}_u - \mathbf{e}_{l_{u}} \|_2^2.
\end{align}
By plugging the KL divergence into Eq.~(\ref{app_eq:vq_2}), we have 
\begin{align}
\mathbb{E}_{\underbrace{Q_{\lambda}(\mathbf{z}^{(q)}_{1:U} \mid \mathbf{x}_{1:T}, \mathbf{y}_{1:U})}_{\text{indicator function}}}\Big\{\mathrm{KL}\big[{Q_{\psi}(\mathbf{{z}}^{(r)}_{1:U} \mid \mathbf{z}^{(q)}_{1:U}, \mathbf{x}_{1:T}, \mathbf{y}_{1:U})} \| {p(\mathbf{{z}}^{(r)}_{1:U} \mid \mathbf{z}^{(q)}_{1:U}, \mathbf{y}_{1:U})}\big]\Big\}   = \sum_{u=1}^{U} \frac{1}{2\sigma^2} \| \mathbf{d}_u - \mathbf{e}_{l_{u}} \|_2^2 \label{app_eq:elbo_vq_2}.
\end{align}

\subsection{In summary}

With Eqs.~(\ref{app_eq:prior-dictionary-learning}) and (\ref{app_eq:elbo_vq_2}), we get the final form for the ELBO:
\begin{align}
\text{ELBO}(\mathbf{x}_{1:T}, \mathbf{y}_{1:U})= & \mathbb{E}_{Q_{\lambda}(\mathbf{z}^{(q)}_{1:U})}[\log p_\theta(\mathbf{x}_{1:T} \mid \mathbf{z}^{(q)}_{1:U}, \mathbf{y}_{1:U})] \label{app_eq:loss_1},\\
& - \sum_{u=1}^U \Big\{ \|\mathbf{c}_u - \mathbf{e}_{l_u}\|_2^2 + \log \sum_{k=1}^K \exp({-\|\mathbf{c}_u - \mathbf{e}_k\|_2^2})\Big\} \label{app_eq:loss_2}, \\
&- \sum_{u=1}^{U} \frac{1}{2\sigma^2} \| \mathbf{d}_u -  \mathbf{e}_{l_{u}}\|_2^2  \label{app_eq:loss_3},
\end{align}
where $\mathbf{d}_u = \mathrm{LatentPredictor}_\psi(\mathbf{z}^{(q)}_{u-1}, \mathbf{\bar{x}}_u, y_u)$ and $\mathbf{c}_u = \mathrm{LatentPredictor}_\phi(\mathbf{z}^{(q)}_{u-1}, y_u)$. 

Our derivation is similar to \citeapp{henter2018deep2}. However, we further take into account the condition $y_{1:U}$ and define a parametric form for the prior in Eq.~(\ref{app_eqn:p_z}) rather than assuming it to be uniform. This parametric prior leads to the loss in Eq.~(\ref{app_eq:loss_2}), which is not included in unconditional models. We further assume a Gaussian for $p(\mathbf{z}^{(r)}_u \mid \mathbf{z}^{(q)}_{u-1}, \mathbf{y}_{1:U}, \mathbf{x}_{1:T}) = \mathcal{F}(\mathbf{z}^{(r)}_{u} -  \mathbf{d}_u)$, rather than assuming it to be a Dirac delta function. 

Note how Eqs.~(\ref{app_eq:loss_1}) and (\ref{app_eq:loss_3}) are similar to the original training criteria of VQ-VAE (with $\beta=1$). Our derivation can thus be used to interpret conditional VQ-VAE.


\subsection{Alternative to Gaussian quantization noise}
As an alternative to the procedure in Section~\ref{app_seq:app_vq}, we may also define the vector quantization part as
\begin{align}
p(\mathbf{{z}}^{(r)}_{1:U} \mid \mathbf{z}^{(q)}_{1:U}, \mathbf{y}_{1:U}) = \prod_{u=1}^{U}p(\mathbf{{z}}^{(r)}_{u} \mid \mathbf{z}^{(q)}_{u})=\prod_{u=1}^{U} Z 
\frac{\mathcal{N}(\mathbf{{z}}^{(r)}_{u}; \mathbf{z}^{(q)}_{u}, \sigma^2\mathbf{I})}{\frac{1}{K}\sum_{k=1}^{K}\mathcal{N}(\mathbf{{z}}^{(r)}_{u}; \mathbf{e}_{k}, \sigma^2\mathbf{I})}, 
\end{align}
where $Z$ is a scalar to normalize the PDF so that $\int p(\mathbf{{z}}^{(r)}_{1:U} \mid \mathbf{z}^{(q)}_{1:U}, \mathbf{y}_{1:U})d\mathbf{{z}}^{(r)}_{1:U} = 1$.  

Then, the KL divergence can be computed as 
\begin{align}
&\mathbb{E}_{\underbrace{Q_{\lambda}(\mathbf{z}^{(q)}_{1:U} \mid \mathbf{x}_{1:T}, \mathbf{y}_{1:U})}_{\text{indicator function}}}\Big\{\mathrm{KL}\big[{Q_{\psi}(\mathbf{{z}}^{(r)}_{1:U} \mid \mathbf{z}^{(q)}_{1:U}, \mathbf{x}_{1:T}, \mathbf{y}_{1:U})} \| {p(\mathbf{{z}}^{(r)}_{1:U} \mid \mathbf{z}^{(q)}_{1:U}, \mathbf{y}_{1:U})}\big]\Big\},   \\
= & \sum_{u=1}^{U} \mathrm{KL}\Big[ \mathcal{N}(\mathbf{{z}}^{(r)}_{u}; \mathbf{d}_u, \sigma^2\mathbf{I}) \|  
\frac{\mathcal{N}(\mathbf{{z}}^{(r)}_{u}; \mathbf{z}^{(q)}_{u}=\mathbf{e}_{l_u}, \sigma^2\mathbf{I})}{\frac{1}{K}\sum_{k=1}^{K}\mathcal{N}(\mathbf{{z}}^{(r)}_{u}; \mathbf{e}_{k}, \sigma^2\mathbf{I})} Z\Big],\\
= & \sum_{u=1}^{U} \Big\{\mathrm{KL}\big[ \mathcal{N}(\mathbf{{z}}^{(r)}_{u}; \mathbf{d}_u, \sigma^2\mathbf{I}) \|  \mathcal{N}(\mathbf{{z}}^{(r)}_{u}; \mathbf{z}^{(q)}_{u}=\mathbf{e}_{l_u}, \sigma^2\mathbf{I}) \big] -\mathrm{KL}\Big[ \mathcal{N}(\mathbf{{z}}^{(r)}_{u}; \mathbf{d}_u, \sigma^2\mathbf{I}) \|
{\frac{1}{ZK}\sum_{k=1}^{K}\mathcal{N}(\mathbf{{z}}^{(r)}_{u}; \mathbf{e}_{k}, \sigma^2\mathbf{I})} \Big] \Big\},\\
= & \sum_{u=1}^{U} \Big\{\frac{1}{2\sigma^2} \| \mathbf{d}_u - \mathbf{e}_{l_{u}} \|_2^2 -\mathrm{KL}\Big[ \mathcal{N}(\mathbf{{z}}^{(r)}_{u}; \mathbf{d}_u, \sigma^2\mathbf{I}) \|
{\frac{1}{ZK}\sum_{k=1}^{K}\mathcal{N}(\mathbf{{z}}^{(r)}_{u}; \mathbf{e}_{k}, \sigma^2\mathbf{I})} \Big]\Big\}.
\label{app_eq:vq_alternative}
\end{align}

For the KL divergence between Gaussian and the mixture of Gaussian, there is no closed form. If we use approximation (variational approximation in \citeapp{durrieu2012lower}), we get
\begin{align}
\mathrm{KL}\Big[ \mathcal{N}(\mathbf{{z}}^{(r)}_{u}; \mathbf{d}_u, \sigma^2\mathbf{I}) \| \frac{1}{K}{\sum_{k=1}^{K}\mathcal{N}(\mathbf{{z}}^{(r)}_{u}; \mathbf{e}_{k}, \sigma^2\mathbf{I})} \Big] \approx  -\log \frac{1}{K}\sum_{k=1}^{K}\exp(-\frac{1}{2\sigma^2} \| \mathbf{d}_u - \mathbf{e}_{k} \|_2^2) \label{app_eq:kl_mixture}.
\end{align}
With Eqs.~(\ref{app_eq:kl_mixture}) and (\ref{app_eq:vq_alternative}), we have
\begin{align}
&\mathbb{E}_{\underbrace{Q_{\lambda}(\mathbf{z}^{(q)}_{1:U} \mid \mathbf{x}_{1:T}, \mathbf{y}_{1:U})}_{\text{indicator function}}}\Big\{\mathrm{KL}\big[{Q_{\psi}(\mathbf{{z}}^{(r)}_{1:U} \mid \mathbf{z}^{(q)}_{1:U}, \mathbf{x}_{1:T}, \mathbf{y}_{1:U})} \| {p(\mathbf{{z}}^{(r)}_{1:U} \mid \mathbf{z}^{(q)}_{1:U}, \mathbf{y}_{1:U})}\big]\Big\},   \\
\approx & \sum_{u=1}^{U} \Big\{\frac{1}{2\sigma^2} \| \mathbf{d}_u - \mathbf{e}_{l_{u}} \|_2^2 + \log \sum_{k=1}^{K}\exp(-\frac{1}{2\sigma^2} \| \mathbf{d}_u - \mathbf{e}_{k} \|_2^2)\Big\}.
\label{app_eq:vq_alternative_full}
\end{align}
Here we ignore the constant term on $\frac{1}{ZK}$. While it is also possible to use  Eq.~(\ref{app_eq:vq_alternative_full}) for the ELBO, in this paper we use the form in Eq.~(\ref{app_eq:elbo_vq_2}).

\section{Sampling of duration with CTC}


Sampled duration $\mathbf{l}_{1:U}$ from $Q_\lambda(\mathbf{z}^{(q)}_{1:U}\mid \mathbf{x}_{1:T}, \mathbf{y}_{1:U})$ must satisfy two constraints:  $1 \le l_u \le K$ and $T = \sum_{u=1}^U l_u$. Furthermore, the sampled $\mathbf{l}_{1:U}$ should be reasonably accurate so that we do not have to draw many samples during model training. Note that the sampled duration is used in  $\mathrm{Aggregate}(\mathbf{x}_{1:T}, \mathbf{z}^{(q)}_{1:U}=\mathbf{l}_{1:U})$ and $\mathrm{Upsample}(\mathbf{y}_{1:U}, \mathbf{z}^{(q)}_{1:U}=\mathbf{l}_{1:U})$ to align the acoustic features $\mathbf{x}_{1:T}$ with linguistic feature sequence $\mathbf{y}_{1:U}$.

\begin{figure}[!t]
\centering
\begin{subfigure}[b]{0.4\textwidth}
\includegraphics[width=1.0\textwidth]{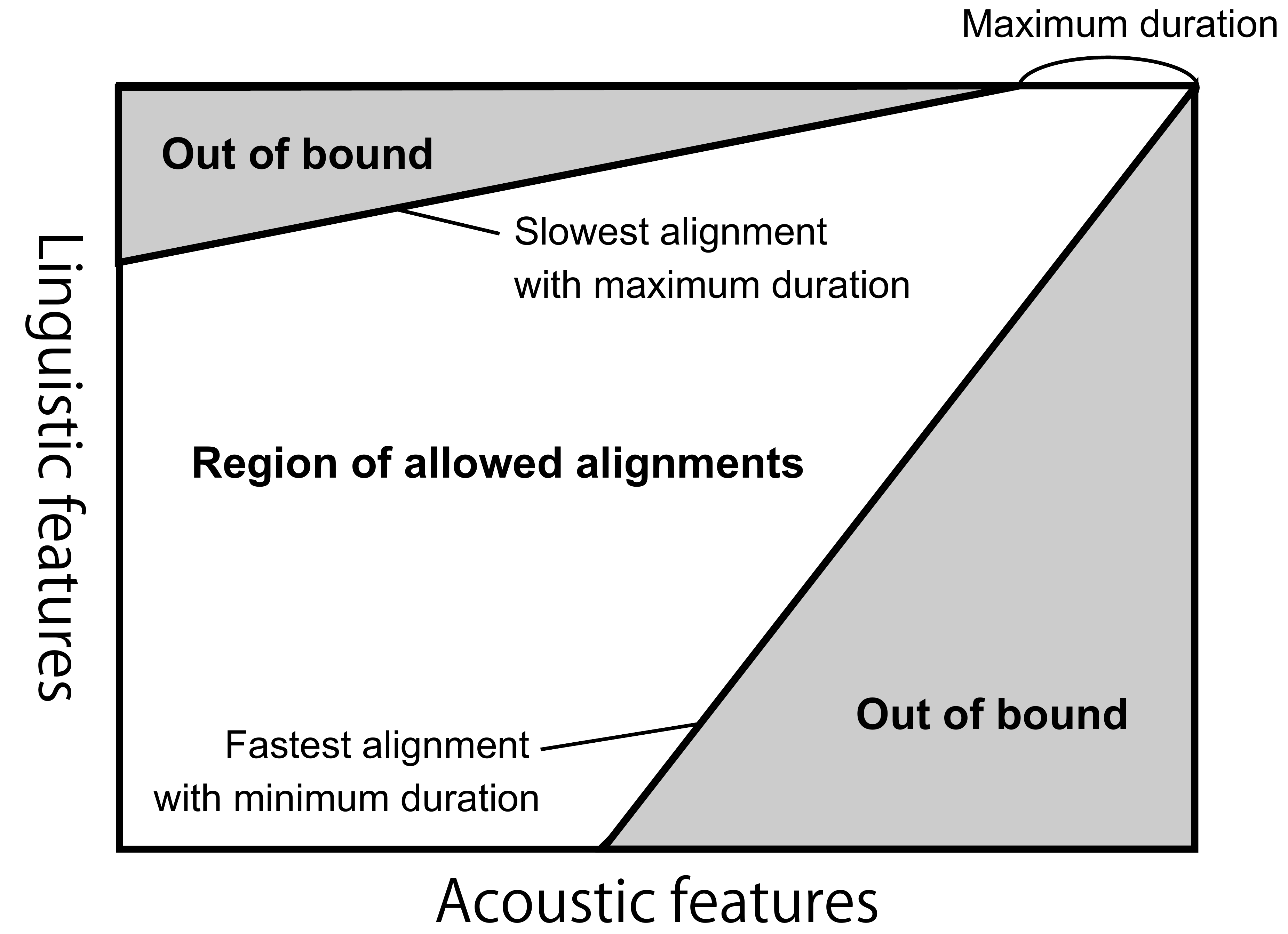}
\caption{}
\label{app_fig:search-condition}
\end{subfigure}
\begin{subfigure}[b]{0.4\textwidth}
\includegraphics[width=1.0\textwidth]{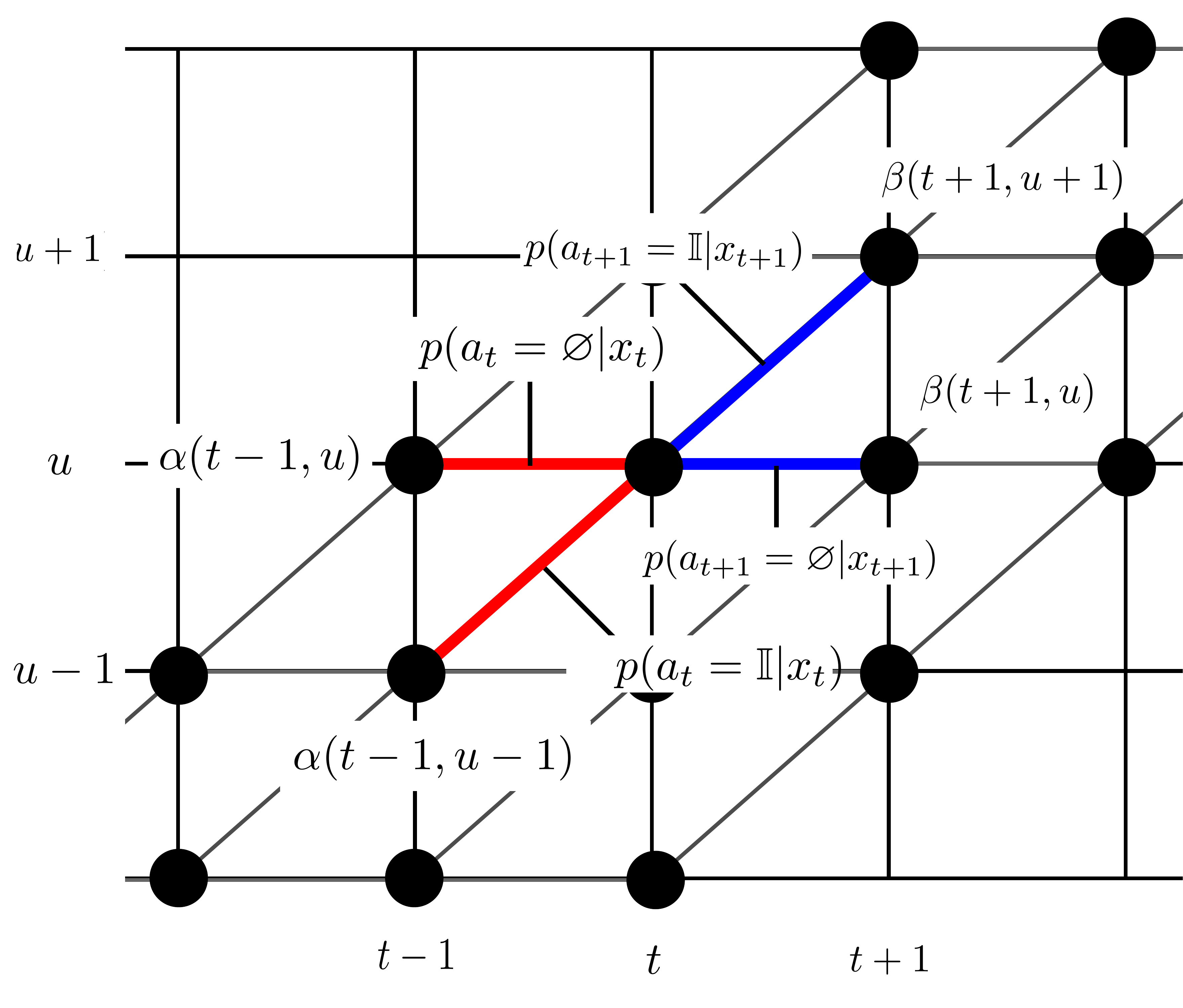}
\caption{}
\label{app_fig:forward-backward}
\end{subfigure}
\caption{(a) Example trellis of CTC-based recognition model and (b) constraints on possible path.}
\end{figure}

The above requirements motivate us to implement $Q_\lambda(\mathbf{z}^{(q)}_{1:U}\mid \mathbf{x}_{1:T}, \mathbf{y}_{1:U})$ with the help of a recognition model $P_\lambda(\mathbf{y}_{1:U} \mid \mathbf{x}_{1:T})$ based on connectionist temporal classification (CTC) \citeapp{Graves2006CTC}. 
Since CTC parameterizes the alignment as a trellis  (see Fig.~\ref{app_fig:forward-backward}), it is straightforward to convert the monotonic CTC alignment into the duration $\mathbf{z}_{1:U}$. Furthermore, the constraint $T = \sum_{u=1}^U z_u$ can be satisfied by only considering the CTC alignments that start from $(t,u) = (1,1)$ and end at $(t,u) = (T,U)$. Alignment that does not satisfy $1 \le z_u \le K$ can also be directly excluded from the trellis, as Fig.~\ref{app_fig:search-condition} illustrates. 
Last but not least, if $P_\lambda(\mathbf{y}_{1:U} \mid \mathbf{x}_{1:T})$ is well trained, we can select the alignment 
$\mathbf{a^\ast}_{1:T}$ that maximizes  $P_\lambda(\mathbf{a}_{1:T} \mid \mathbf{x}_{1:T}, \mathbf{y}_{1:U})$, and $\mathbf{z}_{1:U}$ derived from $\mathbf{a^\ast}_{1:T}$ is expected to be sufficiently accurate.

%

To recap, CTC predicts linguistic features $\mathbf{y}_{1:U}$ from acoustic features $\mathbf{x}_{1:T}$ by marginalizing all possible alignments $\mathbf{a}_{1:T}$. 
The monotonic alignment is represented in alignment transition variable $a_t \in \{\varnothing, \mathbb{I}\}$, where the blank symbol $\varnothing$ means keeping the current linguistic label position, and the shift symbol $\mathbb{I}$ means transition to the next linguistic label position\footnote{Because we are interested in alignment rather than output labels, instead of including blank label $\varnothing$ to output symbols as one class, we separate conditional probability at each time step into alignment transition probability $P_\lambda(a_t|x_t)$ and output probability $P_\lambda(\bar{y}_t|x_t)$.}. Accordingly, the probability of observing $\mathbf{y}_{1:U}$ given $\mathbf{x}_{1:U}$ is defined as
\begin{align}
P_\lambda(\mathbf{y}_{1:U} \mid \mathbf{x}_{1:T}) = \sum_{\forall\mathbf{a}}P_\lambda(\mathbf{y}_{1:U}, \mathbf{a}_{1:T} \mid \mathbf{x}_{1:T}) &\approx \sum_{\forall\mathbf{a}_{1:T}}\prod_{t=1}^T P_\lambda(a_t|x_t)P_\lambda(\hat{y}_t|x_t) \label{app_eq:ctc},\\
\{\hat{y}_1 \cdots, \hat{y}_T\} &= \mathrm{Upsample}(\mathbf{y}_{1:U}, \mathrm{AlignmentToDuration}(\mathbf{a}_{1:T}))\\
\mathbf{z}_{1:U} = \{z_1, \cdots, z_U\} &= \mathrm{AlignmentToDuration}(\mathbf{a}_{1:T}) = \left\{z_u \middle| \sum_{t \in \mathcal{T}(z_{1:u})} \delta(a_t = \varnothing) + 1\right\}.
\end{align}

During training, the likelihood in Eq.~(\ref{app_eq:ctc}) can be computed efficiently with a forward-backward algorithm
(as shown in Fig. \ref{app_fig:forward-backward}) or written as
\begin{align}
P_\lambda(\mathbf{y}_{1:U} \mid \mathbf{x}_{1:T}) &= \sum_{t=1}^T\sum_{u=1}^U\alpha(t, u)\beta(t, u),\\
\alpha(t, u) &= \alpha(t - 1, u)p(a_t=\varnothing|x_t)p(\hat{y}_t|x_t) + \alpha(t - 1, u - 1)p(a_t=\mathbb{I}|x_t)p(\hat{y}_t|x_t)\\
\beta(t, u)  &= \beta(t + 1, u)p(a_{t+1}=\varnothing|x_{t+1})p(\hat{y}_t|x_{t+1}) + \beta(t + 1, u + 1)p(a_{t+1}=\mathbb{I}|x_{t+1})p(\hat{y}_{t + 1}|x_{t + 1}).
\end{align}
This recognition model is jointly trained with other components of the proposed model, as explained in the next section.



To sample a good alignment from the CTC model, note that the $\mathbf{a}^\ast_{1:T}$ that maximizes $P_\lambda(\mathbf{a}_{1:T} \mid \mathbf{x}_{1:T}, \mathbf{y}_{1:U})$
also maximizes the joint probability $P_\lambda(\mathbf{y}_{1:U}, \mathbf{a}_{1:T} \mid \mathbf{x}_{1:T})$. This can be shown by
\begin{align}
\mathbf{a^\ast}_{1:T} 
&= \arg\max_{\mathbf{a}_{1:T}}P_\lambda(\mathbf{a}_{1:T} \mid \mathbf{x}_{1:T}, \mathbf{y}_{1:U}),\nonumber\\
&= \arg\max_{\mathbf{a}_{1:T}}P_\lambda(\mathbf{y}_{1:U} \mid \mathbf{x}_{1:T})P_\lambda(\mathbf{a}_{1:T} \mid \mathbf{x}_{1:T}, \mathbf{y}_{1:U}),\nonumber\\
&= \arg\max_{\mathbf{a}_{1:T}}P_\lambda(\mathbf{y}_{1:U}, \mathbf{a}_{1:T} \mid \mathbf{x}_{1:T}),
\end{align}
where $P_\lambda(\mathbf{y}_{1:U} \mid \mathbf{x}_{1:T})$ is constant against $\mathbf{a}_{1:T}$. Since the recognition model can optimize the joint probability for all possible alignments by marginalization $P_\lambda(\mathbf{y}_{1:U} \mid \mathbf{x}_{1:T}) = \sum_{\forall\mathbf{a}}P_\lambda(\mathbf{y}_{1:U}, \mathbf{a}_{1:T} \mid \mathbf{x}_{1:T})$, the $\mathbf{a}^\ast_{1:T}$ that maximizes $P_\lambda(\mathbf{a}_{1:T} \mid \mathbf{x}_{1:T}, \mathbf{y}_{1:U})$, or equivalently $P_\lambda(\mathbf{y}_{1:U}, \mathbf{a}_{1:T} \mid \mathbf{x}_{1:T})$, is expected to be sufficiently accurate for evaluating the ELBO of the proposed model.

Accordingly, we use the following criterion to acquire $\mathbf{a}^\ast_{1:T}$ and then convert it into the duration sequence $\mathbf{z}_{1:T}$:
\begin{align}
\mathbf{a}^\ast_{1:T} &= \arg\max_{\mathbf{a}_{1:T}}P_\lambda(\mathbf{y}_{1:U}, \mathbf{a}_{1:T} \mid \mathbf{x}_{1:T}) \label{app_eq:search_optimal},\\
\mathbf{l}_{1:U} &= \mathrm{AlignmentToDuration}(\mathbf{a^\ast}_{1:T}).
\end{align}
While we could use a simple greedy search for Eq.~(\ref{app_eq:search_optimal}), the outcome might be inferior due to the independence assumption assumed by CTC. In practice, we search for the N best duration by beam search. 
The search is conducted on the trellis produced by CTC, where the score of each lattice point $\alpha(t,u)\beta(t,u)$ is computed by using its statistics through the forward-backward algorithm.

\section{Detailed discussion on model architecture}

\begin{figure}[!t]
\centering
\includegraphics[trim=10 100 05 250,clip,width=0.8\linewidth]{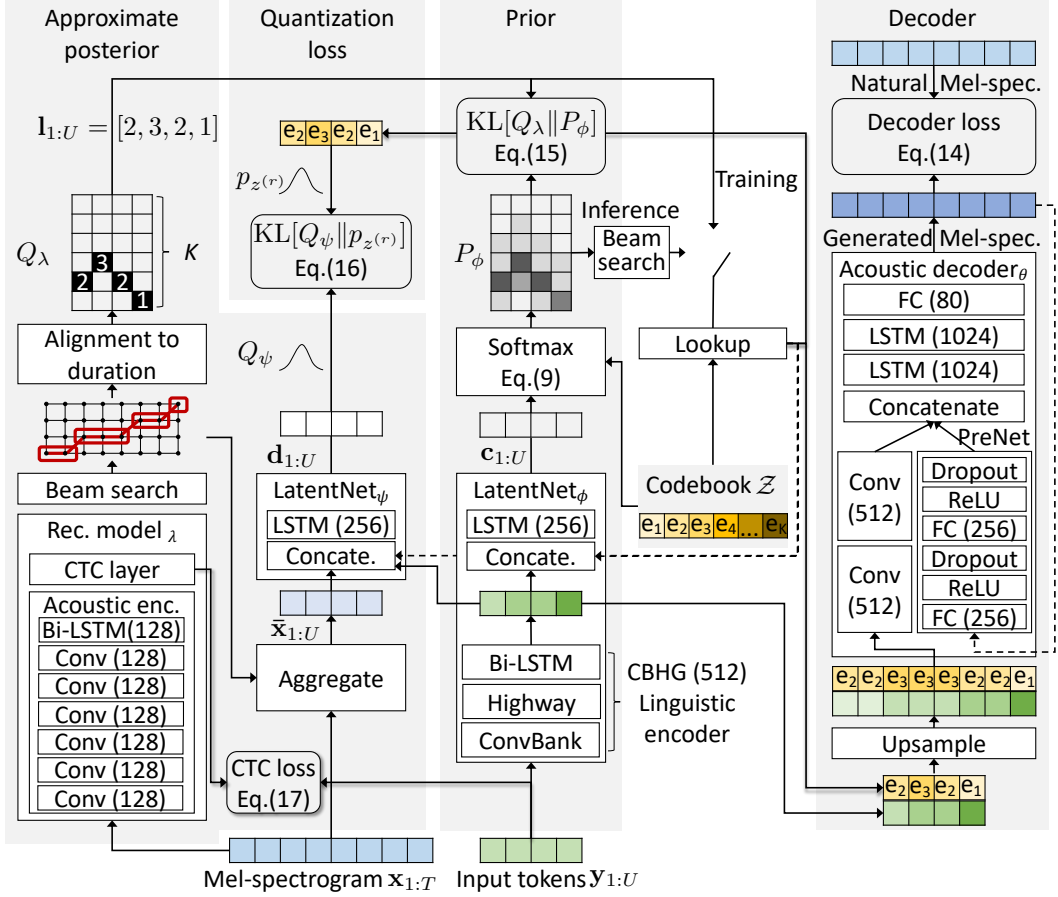}
\caption{Detailed architecture of proposed TTS model. Dashed line denotes feedback loop. Number in bracket denotes neural layer size. FC denotes a fully connected layer. During inference, only prior and decoder are used.}
\label{app_fig:duration-network-structure}
\end{figure}

Figure \ref{app_fig:duration-network-structure} shows the network architecture of our proposed TTS system. The trainable parts include the $\mathrm{LatentNet}_\psi$ and the CTC model in the approximate posterior, $\mathrm{LatentNet}_\phi$ in the prior, the acoustic decoder, and the codebook. The detailed structure is explained as follows.
\begin{itemize}
    \item \textbf{Acoustic encoder in CTC-based recognition model}: It consists of six convolutional layers, each of which has 128 output dimensions and 3 kernels. Its output layer is a bi-directional LSTM layer with 128 dimensions.
    \item \textbf{LatentNet}$_\psi$: It contains an LSTM layer with 256 output dimensions. Its input is the concatenation of the aggregated acoustic features $\mathbf{\bar{x}}_{u}$, the output of the linguistic encoder for the $u$-th token, and the feedback code of the previous token $\mathbf{z}^{(q)}_{u-1} =  \mathbf{e}_{l_{u-1}}$. Note that $\mathrm{LatentNet}_\psi$ sequentially computes $\mathbf{d}_{u}$ from $u=1$ to $u=U$.
    \item \textbf{Linguistic encoder in prior}: It is based on the CBHG encoder in Tacotron. However, the GRU layer is replaced with an LSTM layer with zoneout regularization. The layer size of the CBHG encoder is 512.
    \item \textbf{LatentNet}$_\phi$: It contains an LSTM layer with 256 output dimensions. Its input is the concatenation of the linguistic encoder's output and the feedback code of the previous token $\mathbf{z}^{(q)}_{u-1} =  \mathbf{e}_{l_{u-1}}$. Note that $\mathrm{LatentNet}_\phi$ sequentially computes $\mathbf{c}_{u}$ from $u=1$ to $u=U$.
    \item \textbf{Decoder}: It contains two CNN layers with 512 units and 5 and 7 kernels to process the upsampled linguistic features. The acoustic features fed back from the previous step are transformed with a pre-net, after which they are concatenated with the processed linguistic features and fed to the two LSTM layers with 1024 output dimensions and the output fully connected (FC) layer. The pre-net consists of fully connected layers, ReLU activation functions, and dropout layers.
\end{itemize}

All components are jointly trained. During inference, the prior predicts the duration or codebook indices $\mathbf{l}_{1:U}$, given input $\mathbf{y}_{1:U}$. The retrieved code words $(\mathbf{z}^{(q)}_{l_{1}}, \cdots, \mathbf{z}^{(q)}_{l_{U}})$ and the output of the linguistic encoder are fed to the decoder for acoustic feature generation.

\bibliographystyleapp{IEEEbib}
\bibliographyapp{sample}


\end{document}